\documentclass[CRPHYS,Unicode,manuscript]{cedram}

\usepackage{amsmath}
\usepackage[textwidth=16cm]{geometry}

\DeclareMathAlphabet{\mathcal}{OMS}{cmsy}{m}{n} 

\newcommand{\Patch}[1]{\tilde{#1}}
\newcommand{\Slip}[1]{#1}
\newcommand{\ext}[1]{{#1}^{\rm ext}}
\newcommand{\extxy}[1]{{#1}^{\rm ext}_{\rm xy}}
\newcommand{\av}[1]{\langle {#1} \rangle}
\newcommand{\std}[1]{\textrm{std}({#1})}
\newcommand{\crit}[1]{{#1}^{\rm c}}
\newcommand{\el}[1]{{#1}^{\rm el}}
\newcommand{\pl}[1]{{#1}^{\rm pl}}

\newcommand{\EffShear}{\tau}
\newcommand{\coupling}{\chi}
\newcommand{\orientationSlip}{\theta}
\newcommand{\orientationShear}{\alpha}
\newcommand{\Stress}{\boldsymbol{\Sigma}}
\newcommand{\Strain}{\boldsymbol{\varepsilon}}

\newcommand{\Pressure}{p}

\newcommand{\schmid}{\boldsymbol{M}}
\newcommand{\Stiffness}{\mathbb{C}}
\newcommand{\Eshelby}{\mathbb{S}}
\newcommand{\Identity}{\mathbb{I}}

\newcommand{\preStress}{\Stress^{0}}

\newcommand{\Threshold}{\crit{\EffShear}}
\newcommand{\DeltaThreshold}{\Delta\Threshold}

\newcommand{\SlipThreshold}{\crit{\Slip{\EffShear}}}
\newcommand{\SlipEffShear}{\Slip{\EffShear}}

\newcommand{\extOrientationShear}{\ext{\orientationShear}}

\newcommand{\elStrain}{\el{\Strain}}

\newcommand{\plStrain}{\pl{\Strain}}
\newcommand{\plGamma}{\pl{\gamma}}

\newcommand{\scalarStress}{\Sigma}

\newcommand{\PatchStress}{\Patch{\Stress}}
\newcommand{\PatchScalarStress}{\Patch{\scalarStress}}
\newcommand{\PatchEffStress}{\Patch{\EffShear}}
\newcommand{\critPatchStress}{\crit{\Patch{\Stress}}}

\newcommand{\PatchPressure}{\Patch{\Pressure}}

\newcommand{\PatchThreshold}{\crit{\Patch{\EffShear}}}
\newcommand{\PatchDisThreshold}{\Delta\PatchThreshold}
\newcommand{\PatchDropThreshold}{\delta\PatchThreshold}
\newcommand{\PatchDeltaThreshold}{\crit{\Delta\Patch{\EffShear}}}

\newcommand{\extShearStrain}{\ext{\gamma}}

\newcommand{\extShearStress}{\extxy{\scalarStress}}

\newcommand{\extShearStressIncr}{\Delta \extxy{\scalarStress}}
\newcommand{\extShearStressDrop}{\delta \extxy{\scalarStress}}
\newcommand{\PatchvonMisesStress}{\PatchScalarStress_{\rm vm}}
\newcommand{\PatchDevAngle}{\Patch{\beta}}

\newcommand{\figref}[1]{Fig.~\ref{fig:#1}}
\newcommand{\secref}[1]{Sec.~\ref{sec:#1}}

\newcommand{\supsecref}[1]{\ref{sec:#1}}

\newcommand{\Eqref}[1]{Eq.~\ref{eq:#1}}

\usepackage{xcolor}
\newcommand \David[1] {\bgroup\noindent[\textcolor{blue}{\textbf{David}: #1}]\egroup\ignorespacesafterend}
\newcommand \Sylvain[1] {\bgroup\noindent[\textcolor{red}{\textbf{Sylvain}: #1}]\egroup\ignorespacesafterend}
\newcommand \Stephane[1] {\bgroup\noindent[\textcolor{green}{\textbf{Stephane}: #1}]\egroup\ignorespacesafterend}

\title{Insights from the quantitative calibration of an elasto-plastic model from a Lennard-Jones atomic glass}

\author{\firstname{David} \lastname{Fernández Castellanos}\IsCorresp}
\address{PMMH, CNRS, ESPCI Paris, Universit\'e PSL, Sorbonne Universit\'e, Universit\'e de Paris, 75005 Paris, France}
\email[D. F. C.]{david.fernandez-castellanos@espci.fr}

\author{\firstname{Stéphane} \lastname{Roux}}
\address{Université Paris-Saclay, ENS Paris-Saclay, CNRS, LMT-Laboratoire de Mécanique et Technologie, Université Paris-Saclay, 4 avenue des Sciences, 91192 Gif-sur-Yvette, France}

\author{\firstname{Sylvain} \lastname{Patinet}}
\addressSameAs{1}{PMMH, CNRS, ESPCI Paris, Universit\'e PSL, Sorbonne Universit\'e, Universit\'e de Paris, 75005 Paris, France}



\begin{abstract} 
Quantitative multi-scale modeling of mechanical properties of disordered materials is still an open challenge. Bridging scales requires an intense dialogue between physics and mechanics to keep track of the complexity of the mechanisms at play, especially when passing from a discrete atomistic description to a continuous one. Here, we compare the macroscopic and the local plastic behavior of a model amorphous solid based on two radically different numerical descriptions. On the one hand, we simulate glass samples by atomistic simulations. On the other, we implement a mesoscale elasto-plastic model based on a solid-mechanics description. The latter is extended to consider the anisotropy of the yield surface via statistically distributed local and discrete weak planes on which shear transformations can be activated. To make the comparison as quantitative as possible, we consider the simple case of a quasistatically driven two-dimensional system in the stationary flow state and compare mechanical observables measured on both models over the same length scales. To this end, we first calibrate the macroscale behavior of the elasto-plastic model based on molecular static simulations. We show that the macroscale mechanical response, including its fluctuations, can be quantitatively recovered for a range of elasto-plastic mesoscale parameters. Using a newly developed method that makes it possible to probe the local yield stresses in atomistic simulations,  we calibrate the local mechanical response of the elasto-plastic model at different coarse-graining scales. In this case, the calibration shows a qualitative agreement only for an optimized subset of mesoscale parameters and for sufficiently coarse probing length scales. This calibration allows us to establish a length scale for the mesoscopic elements that corresponds to an upper bound of the shear transformation size, a key physical parameter in elasto-plastic models. We find that certain properties naturally emerge from the elasto-plastic model, such as accurate correlations between external stress fluctuations or between local yield stresses and local stress drops. In particular, we show that the elasto-plastic model reproduces the Bauschinger effect, namely the plasticity-induced anisotropy in the macroscale stress-strain response. We discuss the successes and failures of our approach, the impact of different model ingredients and propose future research directions for quantitative multi-scale models of amorphous plasticity.
\end{abstract}

\begin{document}

\maketitle

\selectlanguage{english}

\section{Introduction}

Multi-scale models of plastic deformation aim to gain a qualitative and quantitative understanding of the relation between microstructural properties and the dynamics of plastic activity. Ultimately, such models can help design new microstructures tailored to meet the demands of specific engineering applications and industries~\cite{van_der_Giessen_2020,McDowell2007}. An essential ingredient for a successful multi-scale approach is establishing links between atomistic and macroscale continuum descriptions in a physically grounded manner. In this perspective, an intermediate mesoscale description is a desirable conceptual step, as illustrated, for instance, in crack propagation~\cite{patinet_cracks_2014} and crystal plasticity~\cite{devincre_physically_2015}. Mesoscale approaches rely on a statistical description of the relevant phenomena, enhancing our understanding of the collective processes at play. By doing so, it overcomes the limitations of atomistic and continuum approaches. Namely, atomistic models are limited in terms of system sizes and time scales. On the other hand, continuum approaches have difficulties in capturing complex heterogeneous flow patterns or spatio-temporal correlations in plastic activity~\cite{Rodney2011}. Specifically, mesoscale models circumvent these limitations by replacing the quasi-infinite microscopic degrees of freedom at the atomistic scale with more manageable, discrete, and coarser ones, based on a continuum description~\cite{nicolas_deformation_2018}.

In amorphous materials, plastic deformation occurs at the lowest scales via atomic rearrangements leading to localized shear transformations (STs)~\cite{Argon1979,Tanguy2006,Lerner2009} and, in some cases, to a permanent dilation or contraction associated with local changes of free volume~\cite{molnar_densification_2016}. The properties of STs have been studied in detail and present large distributions of activation energy barriers~\cite{Rodney2009}, directions~\cite{Nicolas2018}, sizes~\cite{Albaret2016} and shapes~\cite{Tanguy2006} and lead to a redistribution of elastic stresses within the system~\cite{talamali_path-independent_2008}. Thus, in these materials, plastic deformation is made up of "quanta" corresponding to local topological changes, which, unlike dislocations~\cite{patinet_atomic-scale_2011}, are localized in space and time~\cite{falk_dynamics_1998}. 

This view of plasticity has been successfully implemented in discrete elasto-plastic models based on a solid-mechanics description~\cite{bulatov_stochastic_1994-1}. The elasto-plastic approach considers a discrete-continuum medium endowed with stochastic local evolution rules for plastic activity and structural properties~\cite{jagla_shear_2010,Nicolas2014,karimi2017,BudrikisNatCom,Nicolas2018,Jagla2020}. One advantage of elasto-plastic models is that they rely on physically meaningful quantities that are easy to interpret. At the same time, they can qualitatively reproduce the phenomenology associated with amorphous plasticity in many conditions such as, e.g., under athermal quasistatic shear~\cite{Talamali2012,tyukodi_depinning_2016,BudrikisNatCom,karimi2017,Jagla2020}, at a finite strain-rate~\cite{liu_driving_rate_2016,Homer2009}, under creep loading~\cite{Castellanos2019} or near mechanical failure~\cite{Tuszes2017,Castellanos2018}. This simplicity and versatility can help establish links between statistical physics and engineering formulations of plasticity. 

Nonetheless, the lack of well-defined plastic deformation units in amorphous solids has hindered the development of multi-scale strategies. Elasto-plastic models have thus remained widely phenomenological or, at best, based on educated guesses. In the shear transformation zone theory framework~\cite{Langer2008}, Hinkle and Falk have for instance employed the potential energies of the atomistic model coarse-grained at different length scales as a surrogate of the effective temperature of the continuum-level field description~\cite{hinkle_coarse_2017}. However, elasto-plastic mesoscale parameters need to be material specific and realistic to make quantitatively accurate predictions. For this reason, there have been many attempts in recent years to measure mesoscale parameters from atomistic simulations~\cite{Schuh2003,Rodney2009,puosi_probing_2015,Nicolas2015,Albaret2016,Boioli2017,Nicolas2018}.

\begin{figure}[h]
	\centering
	\includegraphics[width=13cm, keepaspectratio]{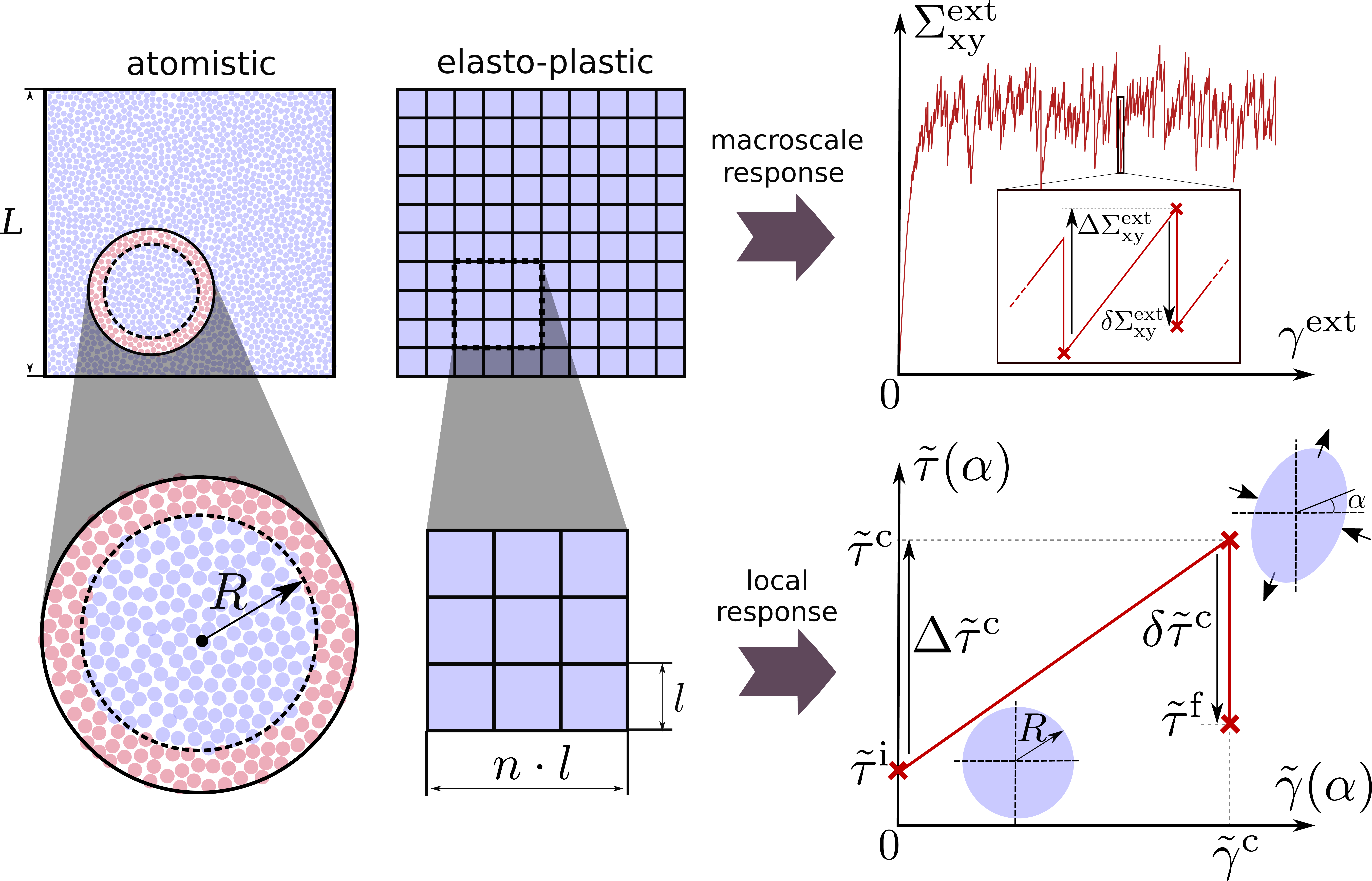} 
	\caption{Multi-scale approach: the macroscopic (top) and local (bottom) quasi-static mechanical responses are investigated in both elasto-plastic and atomistic bi-dimensional models. To measure the local yield stress, patches of equal area are isolated from the rest of the system.}
	\label{fig:multiscale_strategy}
\end{figure}

A fundamental ingredient of elasto-plastic models corresponds to a local measure of shear susceptibility, able to predict in principle the plastic activity~\cite{Tsamados2009,Tanguy2010,Ding2014,Jack2014,Cubuk2015,Patinet2016,Cubuk2017,Wei2019,xu_atomic_2020,Bapst2020}. Among the proposed alternatives, the local yield stress exhibits an excellent correlation with plastic event locations in comparison with other indicators~\cite{Richard2020}. Direct measurements of local yield stress in atomistic samples have become recently possible using a method ~\cite{Patinet2016} based on straining subsets of atoms, as illustrated in \figref{multiscale_strategy}. The method can characterize the local resistance to irreversible deformation, which can be interpreted as a mesoscale generalization of the macroscale yield stress signaling the onset of plastic deformation. Therefore, the local yield stress is well suited for elasto-plastic models, rooted in the framework of solid mechanics. Moreover, the method makes it straightforward to gather statistics of diverse local properties that traditionally have been challenging to measure, such as rearrangement amplitudes and directions~\cite{Barbot2018}, structural evolution~\cite{Barbot2020} or local yield surface anisotropy~\cite{Patinet2020}. However, until recently~\cite{liu_elasto-plastic_2020}, these promising properties have not yet been explored to tentatively fill the gaps necessary for quantitatively accurate elasto-plastic models.

Filling such gap is the aim of the present work in which we compare, as quantitatively as possible, atomistic and elasto-plastic models both at the macro and local scales by leveraging the advantages of local yield stress measurements. We prepare atomistic glass samples driven in the athermal quasistatic limit until the stationary flow state. The local mechanical properties of the as-quenched glasses and the stationary states are measured with the local yield stress method described in~\cite{Patinet2016,Barbot2018} at different length scales and for different shear orientations. On the other hand, we set up a mesoscale elasto-plastic model with statistically distributed structural properties and anisotropic local yield functions and mimic the atomistic glass preparation and loading protocol. Contrary to~\cite{liu_elasto-plastic_2020}, we focus on the athermal quasistatic regime and consider a generalization of the local yield stress method of~\cite{Patinet2016,Barbot2018} for different length scales. Motivated by local atomistic observations showing anisotropic yield surfaces and non-normal plastic flow~\cite{Barbot2018}, we make a step forward by modeling the possibility of plastic rearrangements according to a set of various local discrete weak planes for each element of the mesoscale elasto-plastic model. We investigate the consequences of this feature in terms of model calibration and emergent plastic anisotropy upon plastic deformation. The comparison of behaviors at local scales is carried out by measuring the response of different subsets of the elasto-plastic matrix by mimicking the atomistic procedure (see \figref{multiscale_strategy}). As a result, we do not establish a hierarchical dependence between models, in the sense that atomistic data does not feed the elasto-plastic model. Instead, we consider atomistic and elasto-plastic descriptions on an equal footing and calibrate the elasto-plastic model by requiring the same mechanical properties on both models when measured on the same length scale. In this way, it becomes possible to not only calibrate the model but to assess its predictions at different scales simultaneously, which allows us to carefully analyze agreements and discrepancies with the reference atomistic glass. 

The article is organized as follows: in the second and third sections, we introduce the atomistic and elasto-plastic models and methods employed to characterize both the global and local mechanical response of glasses. Section 4 deals with the calibration of the elasto-plastic model and the comparison of atomistic and mesoscale approaches at both  macroscopic and local scales. The various emergent phenomenologies obtained from the elasto-plastic model are presented in section 5. In section 6, we discuss the successes and failures of our approach, different key ingredients, and suggest future research directions for quantitative multi-scale models of amorphous plasticity.

\section{Atomistic model and methods}
\label{sec:atomistic_model_and_methods}

\subsection{Preparation and loading protocols}
\label{sec:atomistic_system}

We use the two-dimensional binary atomic system studied extensively in~\cite{Patinet2016,Barbot2018}. This system has been widely used as a model glass for its good glass-formability and to study the plasticity of glasses from the atomic scale~\cite{falk_dynamics_1998}. It makes it possible to reproduce the main features of the phenomenology observed in the mechanical response of amorphous solids while remaining as simple as possible. The composition is chosen so that the ratio between the number of large (L) and small (S) particles is equal to $N_{\rm L}/N=(1+\sqrt{5})/4$. The atoms, all with a mass $m=1$, interact via a Lennard-Jones (LJ) interatomic potential of parameters $\sigma_{\rm AB}$ and $\epsilon_{\rm AB}$, where AB corresponds to the interacting species. The inter-species parameter of the potential defines the units of length $\sigma_{\rm SL}$, energy $\epsilon_{\rm SL}$, time $t_{\rm 0}=\sigma_{\rm SL}\sqrt{m/\epsilon_{\rm SL}}$ and stress $\Sigma=\epsilon_{\rm SL}/\sigma_{\rm SL}^2$, used in the rest of the article. For distances between $R_{\rm in}=2$ and the cutoff radius $R_{\rm out}=2.5$, the LJ potential is replaced by a polynomial function to be twice differentiable.

The systems are simulated at a constant density $\rho=1.02$ using periodic boundary conditions. Except for the simulations exploring finite-size effects, where larger systems are employed (see \figref{emerging}b), the systems contain $10^4$ atoms with a linar size equal to $L=98.8045$. The glasses are obtained by instantaneously quenching at constant volume a supercooled liquid at thermodynamic equilibrium from a temperature $T=1.132~T_{\rm g}^{\rm sim}$, where $T_{\rm g}^{\rm sim}=0.31~\epsilon_{\rm SL}/k_{\rm B}$. The quench is followed by a static relaxation to cancel forces via a conjugate gradient method down to machine precision. This preparation protocol, named equilibrated supercooled liquid (ESL) in~\cite{Barbot2018}, has the advantage of producing glass samples which exhibit no stress overshoot when strain is applied and thus without sample-scale shear banding. This is specially convenient for studying the stationary flow regime. 

We consider $100$ independent glass samples deformed by the athermal quasi-static shear (AQS) deformation method~\cite{maloney_universal_2004} in simple shear along the $\textrm{xy}$ direction until a deformation $\extShearStrain=5$. This iterative method consists of deforming the system by small increments of affine strain $\Delta\extShearStrain=10^{-4}$, followed by a static relaxation. The trajectories obtained consist of a series of states at mechanical equilibrium corresponding to the limit of zero temperature and strain rate. In the large strain regime, the typical mechanical response reported in \figref{macro}a is characterized by the fluctuation of the measured stress $\extShearStress$ around a plateau value. As shown schematically in \figref{multiscale_strategy}, it consists of a succession of quasi-linear elastic branches, where the external stress increases by increments $\Delta\extShearStress$, intercepted by stress drops $\delta\extShearStress$ that corresponds to the occurrence of irreversible deformation events dissipating elastic energy. The steady-state, signaled by the convergence of stresses and potential energy, is reached for a strain $\extShearStrain>0.5$.

\subsection{Atomistic implementation of the local shear test}
\label{sec:atomistic_probing_the_local_yield_stresses}

To measure the local mechanical response, we use the recently developed local yield stress method~\cite{puosi_probing_2015,Patinet2016,Barbot2018}, as illustrated in \figref{multiscale_strategy}. This method consists of deforming in pure shear along direction $\orientationShear$ a zone of radius $R$, known as a patch, by applying a purely affine deformation to the surrounding medium. The central area is deformed using the same AQS method as for deforming the overall system. Therefore, plastic events are necessarily triggered within the relaxed zone, which makes it possible to measure the local coarse-grained yield stress $\PatchThreshold$ in several directions independently of other plastic rearrangements. Since glasses are frustrated systems, they are locally decorated with a spatially heterogeneous local stress $\PatchEffStress^{\rm i}$. The residual plastic strength, or distance to threshold $\PatchDisThreshold$, corresponds to the amount of stress added at the coarse-graining scale $R$ to trigger the instability locally, thus $\PatchDisThreshold = \PatchThreshold-\PatchEffStress^{\rm i}$. This method also makes it possible to estimate the relaxation amplitude $\PatchDropThreshold$ by measuring the stress drop associated with the plastic instability.

The sampling of $\PatchEffStress^{\rm i}$, $\PatchThreshold$, $\PatchDisThreshold$, and $\PatchDropThreshold$ is performed in the as-quenched state and the stationary state on a regular square grid with lattice parameter $2.5$. These local quantities are computed for different patch sizes $R$ and orientations $\orientationShear$. We note that the local yield stresses and distances to threshold (stress drops) are expected to be overestimated (underestimated) for too small patch radii $R$ as a consequence of the hard boundary conditions imposed by the frozen matrix of the local yield stress method. A more accurate measure is thus anticipated as $R$ grows. This aspect is discussed further when we compare the results obtained with the elasto-plastic model.

\section{Elasto-plastic model}
\label{sec:elasto-plastic_model}

Elasto-plastic models define the state of a material subvolume in terms of continuum mechanics fields. Here, we discretize the material domain into a 2D regular lattice of mesoscale elasto-plastic quadrilateral elements. Each element corresponds to a material sub-volume where localized atomistic rearrangements can occur. Such rearrangements are represented, at the mesoscale, as localized slip events. This coarse-grained description aims at capturing the essential effects of the rearrangements while neglecting unnecessary details. As illustrated in \figref{multiscale_strategy}, the elements have a length $l$ that is above the typical length scale of the atomistic rearrangements. Below this scale, the elasto-plastic model cannot resolve microscopic details. Consequently, we consider strain fields which are element-wise constant, and associate each mesoscale element with a single elastic and plastic strain tensor, $\elStrain$ and $\plStrain$ respectively.

The elastic strain field $\elStrain$ is the consequence of externally imposed boundary conditions and the presence of the plastic strain field $\plStrain$. We solve the stress equilibrium equation assuming linear elasticity applying the Finite Element Method (FEM) \cite{Homer2009,Sandfeld2015,karimi_role_2016,BudrikisNatCom,Castellanos2018, Castellanos2019}. To this end, we consider a 2D regular grid of quadrilateral finite elements (FEs) with linear shape functions. Each FE exactly matches the shape of a single mesoscale element. By applying the FEM, we obtain the displacement field from where we compute the strain field $\Strain$ as the symmetric gradient. The strain associated with a mesoscale element is defined as the strain averaged within the corresponding FE. The elastic strain is obtained as $\elStrain = \Strain - \plStrain$ and the stress is computed from linear elasticity as $\Stress = \Stiffness:\elStrain + \preStress$, where $\Stiffness$ denotes the rank-4 stiffness tensor and $\preStress$ is the pre-stress present in the system prior to the initiation of the driving protocol. To establish as much contact as possible with the atomistic model, we compute the solution to the stress equilibrium equation under bi-periodic boundary conditions with an externally applied shear strain $\extShearStrain$ along the $\textrm{xy}$ direction. The external shear stress $\extShearStress$ is computed as the average shear stress over all the elasto-plastic elements composing the system.

In the next subsections, we detail the structural properties of the mesoscale elements and their stochastic evolution rules. Plastic activity occurs as a sequence of slip events coupled by stress redistribution competing against structural disorder. The specific choices we make regarding these rules are motivated by measurements on the atomistic systems, against which the elasto-plastic model results will be bench-marked later.

\subsection{Local slip systems}
\label{sec:local_slip_systems}

Plastic activity in glasses proceeds by atomistic rearrangements, which lead, in general, to a permanent localized deformation with a shear component and a hydro-static one, the latter associated with the creation or annihilation of free volume. Nonetheless, following the strong correlation between local shear thresholds and plastic activity~\cite{Barbot2018,Richard2020} as well as the weak correlation between local free volume and softness~\cite{Barbot2020} in the atomistic system considered here, we simplify the description and consider only the shear component.

Even when the tensorial formulation of plasticity is accounted for, the vast majority of the elasto-plastic models assume an isotropic yield criterion to simulate the activation of plastic deformation~\cite{BudrikisNatCom,Castellanos2018,nicolas_deformation_2018}. However, atomistic measurements of the local response to shear of glasses reveal anisotropic yield surfaces and the presence of weak planes~\cite{Barbot2018} with spatially-fluctuating properties. To recreate such a complex response to locally applied shear, we consider that each mesoscale element contains several slip systems. We define each slip system by a plane of normal unit vector $\boldsymbol{n}$ and a direction $\boldsymbol{s}$, contained within the plane. A slip of amplitude $\Delta\plGamma$ is thus represented by the plastic strain increment
\begin{equation}
\label{eq:plastic_increment}
\Delta\plStrain = \Delta\plGamma \schmid,
\end{equation}
with
\begin{equation}
\schmid =  \frac{1}{2} \left( \boldsymbol{s} \otimes \boldsymbol{n} + \boldsymbol{n} \otimes \boldsymbol{s} \right)
\end{equation}
Since we consider a 2D scenario, each slip system contains two possible slip senses $+\boldsymbol{s}$ and $-\boldsymbol{s}$. However, we associate each slip system with a single sense. The reason for this choice is that, in the presence of strong structural disorder, planes that qualify as weak under the action of local shear stress may not qualify as such when the shear direction is reversed~\cite{Barbot2018}. We can write the tensor $\schmid$ in terms of the angle $\orientationSlip$ between the slip plane and the horizontal axis as
\begin{equation}
\schmid(\orientationSlip) =  \frac{1}{2}\begin{pmatrix} -\textrm{sin}2\orientationSlip & \textrm{cos}2\orientationSlip \\ \textrm{cos}2\orientationSlip & \textrm{sin}2\orientationSlip \end{pmatrix}
\end{equation}
with $\orientationSlip \in (-\pi/2,\pi/2]$ due to symmetry. 
The \emph{resolved shear stress} $\SlipEffShear$ on a slip plane is given by
\begin{equation}
\label{eq:resolved_shear_stress}
\SlipEffShear = \schmid(\orientationSlip) : \Stress
\end{equation}
Each slip system has a critical resolved shear stress $\SlipThreshold > 0$, in the following referred to as \emph{slip threshold}. The \emph{distance to threshold} is defined as $\DeltaThreshold=\SlipThreshold-\SlipEffShear$. Whenever a slip system fulfills $\DeltaThreshold<0$, it is denoted as \emph{active}, and a slip event is performed as described later.

\subsection{Structural properties}
\label{sec:distribution_slips}

To set the density of slip systems, we consider that a zone of $l \approx 2$ would contain an average of $4$ atoms and a strict maximum number of slip systems equal to $N=4$. Consequently, a reasonable upper bound for the slip system density is $1$. In our simulations, we set the density of slip systems to $\rho_{\rm s} = 0.8$, below the estimated upper bound (we assess the robustness of our results upon variations of this parameter in Appendix \supsecref{number_slip_systems}). We relate the number $N$ of slip systems to the discretization length scale $l$ as $N = \rho_{\rm s} l^2$.

Both the slip angles and thresholds are statistically distributed to represent structural heterogeneity. The orientation of the local slip systems defined within an element, although statistically distributed, must fulfill a certain constraint. Namely, the element must have a finite critical resolved shear stress $\SlipThreshold(\orientationShear)$, i.e. defined for any shear orientation $\orientationShear$. The simplest way to achieve this is to consider at least four slip systems with $\orientationSlip$, $\orientationSlip + \pi/2$, $\orientationSlip + \pi/4$ and $\orientationSlip + 3\pi/4$ which ensures the fulfillment of the constraint. Due to the a priori lack of privileged orientation, we consider $\orientationSlip$ uniformly distributed in the interval $(-\pi/2, \pi/2]$. We populate each mesoscale element with $N$ multiple of $4$ to apply the procedure defined above.

Slip thresholds are independently drawn, with the only requirement of being positive-definite. For simplicity, we use the Weibull distribution
\begin{equation}
\label{eq:weibull_thresholds}
P(\SlipThreshold | \lambda, k) = \frac{k}{\lambda}\left( \frac{\SlipThreshold}{\lambda} \right)^{k-1}\textrm{exp}\left[ - \left(\frac{\SlipThreshold}{\lambda}\right)^{k} \right],
\end{equation}
where the parameter $\lambda$ defines the scale and the exponent $k$ the shape of the distribution.

\subsection{Performing a slip event}
\label{sec:slip_event}

When a slip event occurs within an element, we update the plastic strain field $\plStrain$ by adding a tensorial increment $\Delta\plStrain$ defined by \Eqref{plastic_increment}, which is homogeneous through the element and zero everywhere else. We consider the amplitude $\Delta\plGamma$ of the strain increment statistically distributed to represent the effects of the heterogeneous microstructure on an initiated slip event. We impose the constraint of avoiding negative dissipation~\cite{vasoya_2020}. Within our model (Appendix \supsecref{slip_amplitudes}), this constraint is fulfilled if the slip amplitude verifies $\Delta\plGamma < \gamma_{\rm max}$ with
\begin{equation}
\label{eq:max_amplitude}
\gamma_{\rm max}(\SlipEffShear) = \frac{-2 \SlipEffShear }{(\Stiffness:(\Eshelby-\Identity):\schmid):\schmid},
\end{equation}
where $\Eshelby$ is the Eshelby tensor of the elasto-plastic elements~\cite{Eshelby1957}. We consider a bounded distribution in order to explicitly verify the constraint. Specifically, we choose a bounded power-law distribution of the form
\begin{equation}
\label{eq:pdf_stress_drop}
P(\Delta\plGamma | \gamma_{\rm max}, \coupling) = \frac{\coupling}{\gamma_{\rm max}}(1-\frac{\Delta\plGamma}{ \gamma_{\rm max}})^{\coupling-1}
\end{equation}
with $0 < \Delta\plGamma < \gamma_{\rm max}$ and $\coupling > 0$. This distribution has the advantage of having a single free parameter, which allows to vary its shape while limiting the complexity of the model calibration. We note, however, that its specific functional form does not significantly affect the results, as will be shown in \secref{successes_failures_and_key_ingredients}.

After a slip event has occurred in a certain element, we renew the orientations and thresholds of the $N$ slip systems within the element from their respective probability distributions. The value of $N$ remains fixed. This process aims to represent stochastic changes in the local microstructural properties induced by plastic activity.

\subsection{Loading protocol}
\label{sec:loading_protocol}

We implement an athermal quasistatic protocol to represent as faithfully as possible the AQS driving protocol of the atomistic model (\secref{atomistic_system}). In the absence of temperature effects, we consider that whenever one or more slip systems are active, slip events are simultaneously performed in all those systems (\secref{slip_event}). Due to stress redistribution, new slip systems might become active. This process is repeated in a series of relaxation steps until no slip system is active. During this process, the external strain $\extShearStrain$ is kept fixed. In the case where several slip systems within the same mesoscale element simultaneously become active, only the slip system with the lowest distance to threshold $\DeltaThreshold$ is considered active.

When there is no active slip system, we perform discrete external shear strain increments of $\Delta\extShearStrain = 10^{-4}$. Slip events taking place between two external strain increments are known as an \emph{avalanche}. We apply external strain increments until reaching a maximum external strain target of $\extShearStrain=2$.

\subsection{Elasto-plastic implementation of the local shear test}
\label{sec:local_shear_tests}

To reproduce as faithfully as possible the results measured at local scales in atomistic simulations, we mimic the method described in \secref{atomistic_probing_the_local_yield_stresses} with the elasto-plastic model. Elasto-plastic patches are defined respecting the geometrical restrictions imposed by the lattice discretization. Thus, we define quadrilateral patches formed by $n \times n$ contiguous elasto-plastic elements selected from the system (\figref{multiscale_strategy}). We associate to each elasto-plastic patch an effective radius $R = (l/\sqrt{\pi})n$ by considering the radius of a circular patch of equal area. This allows us to compare the results with atomistic patches of a specific radius. 

To measure the mechanical response of a patch, we consider it in isolation from the rest of the system (\figref{multiscale_strategy}, left). The stress $\PatchStress$ of a patch is defined as the average stress of the $n \times n$ elements forming the patch. Before the initiation of a shear test, a patch has an initial stress $\PatchStress^{\rm i}$. We apply to its free boundary a pure shear strain with orientation $\orientationShear$ and amplitude $\Patch{\gamma}$. When one slip system becomes active, the patch has reached a critical stress $\critPatchStress$. At this point, an avalanche of slip events occurs at a constant external strain. Afterward, the patch has a final stress $\PatchStress^{\rm f}$. This process is illustrated in \figref{multiscale_strategy} (right). Except for the non-periodic boundary conditions, elasto-plastic patches are driven, and slip events occur, according to the same rules described for the full system (\secref{elasto-plastic_model}).

We note that, although the patch is led to mechanical instability by raising the stress till $\critPatchStress$, this value corresponds to the total amount of stress at the moment of instability. Thus, it depends only on the patch structural properties. We can use it to establish definitions which closely mimic the atomistic ones. Thus, we define the coarse-grained yield stress for the orientation $\orientationShear$ as $\PatchThreshold = \schmid(\orientationShear) : \critPatchStress$ that is, the resolved shear stress on the shearing plane $\orientationShear$ at the moment of the patch mechanical instability. In order to characterize local stability in a manner analogous to the atomistic method, we define the coarse-grained distance to threshold $\PatchDisThreshold$ as $\PatchDisThreshold = \PatchThreshold - \PatchEffStress^{\rm i}$ where $\PatchEffStress^{\rm i} = \schmid(\orientationShear) : \PatchStress^{\rm i}$. Moreover, we define the coarse-grained stress drop $\PatchDropThreshold$ from the yield stress to the final stress state as $\PatchDropThreshold = \PatchThreshold - \PatchEffStress^{\rm f}$.

We remark the difference between the coarse-grained patch-scale yield stress $\PatchThreshold$ and the slip thresholds $\SlipThreshold$ defined in \secref{local_slip_systems}. The former characterizes the measured resistance to plastic deformation at the coarse-graining scale defined by the patch radius $R$, while the latter refers to the resistance to slip of individual weak planes within the patch, with values drawn from \Eqref{weibull_thresholds}.

\section{Calibration of the elasto-plastic model}
\label{sec:calibration_of_the_elasto_plastic_model}

In this section, we calibrate the mesoscale resolution $l$ and the elasto-plastic parameters $\lambda$, $k$, and $\coupling$ associated with the threshold scale, threshold disorder, and slip event amplitude, respectively. The elasto-plastic model considers isotropic and homogeneous elastic properties. To this end, we use the system-scale effective shear modulus $G=13.2$ and bulk modulus $B=59$ (in LJ units) measured in the atomistic model.

First, we focus our attention on the macroscale behavior in the stationary state, which does not depend on the initial conditions. Macroscale behavior provides us with an initial general understanding of the system dynamics. Moreover, it introduces constraints in the mesoscale parameter space, establishing a strategy to calibrate the local spatially-fluctuating scale-dependent mesoscale properties. We perform the calibration at different element length scales $l$ and establish an optimum range of values.

In addition, we also take advantage of this calibration procedure to generate reasonable quench states by assuming that the thermally-activated structural changes occurring in glass-forming liquids are statistically close to those induced by plastic activity~\cite{bulatov_stochastic_1994-2}. The optimal set of parameters are thus resorted to generating the quench states from a Kinetic Monte Carlo approach where only the temperature needs to be calibrated, as detailed in Appendix \supsecref{supercooled_simulation}.

\subsection{Macroscale behavior}
\label{sec:macro_behavior}

\begin{figure}
	\centering
	\includegraphics[width=12cm, keepaspectratio]{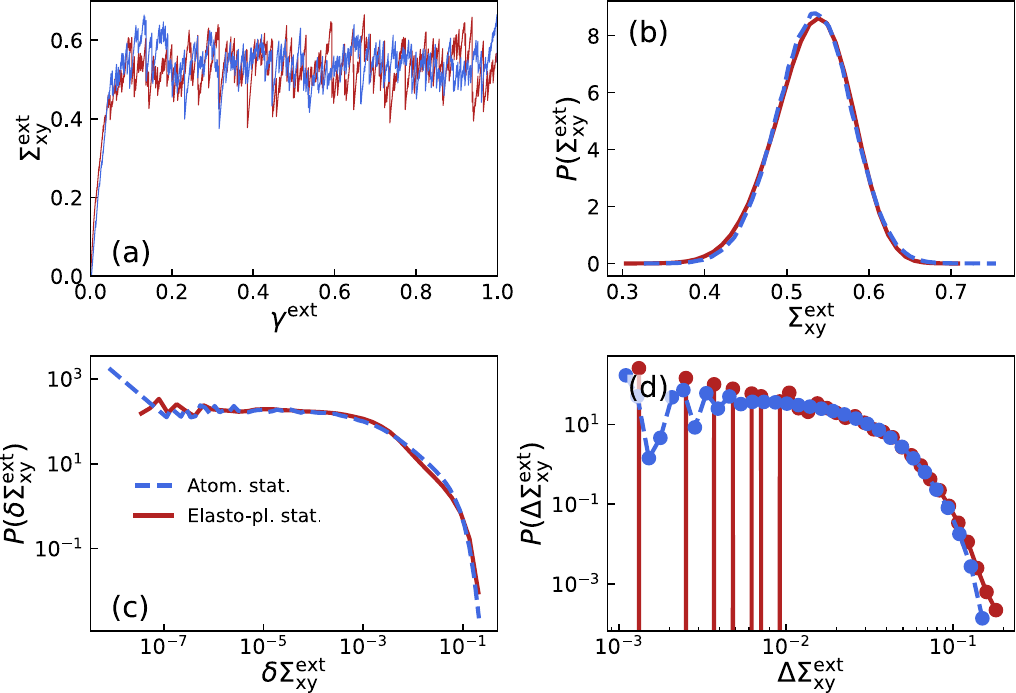} 
	\caption{Comparison of the macroscale behavior of the atomistic (blue) and elasto-plastic (red) models in the stationary regime ($\extShearStrain>0.5$). (a) Single simulation stress-strain curve. Probability densities of: (b) external stress values $\extShearStress$, (c) drops $\extShearStressDrop$ and (d) increments $\extShearStressIncr$ induced by discrete steps of $\Delta \extShearStrain = 10^{-4}$. The results shown correspond to the optimal parameters with $l=6.6$.}
	\label{fig:macro}
\end{figure}

As the externally applied strain $\extShearStrain$ increases, the system exhibits transient behavior related to the initial system configuration. Eventually, the system reaches a stationary regime in which its statistical properties are independent of the external strain (\figref{macro}a). This regime can be identified by a plateau in the external stress $\extShearStress$. For our analysis, we focus on the region $\extShearStrain>0.5$. To quantitatively calibrate the macroscale behavior, we compute the overlap between the elasto-plastic and the atomistic probability densities of external stress values $\extShearStress$, increments $\extShearStressIncr$ induced by the loading mechanism and drops $\extShearStressDrop$ caused by avalanches of plastic activity. Based on the overlap, we define an error function $\mathcal{L}[P]$ for each probability density as 
\begin{equation}
\label{eq:macro_error}
\mathcal{L}[P] = \left[1 - \int_{-\infty}^{+\infty}\textrm{min}\{ P_{\rm EP}(x),\: P_{\rm MD}(x) \}dx \right],
\end{equation}
where $P_{\rm EP}$ and $P_{\rm MD}$ respectively refer to the elasto-plastic and atomistic versions of the probability densities. When the densities perfectly overlap, $\mathcal{L}[P]=0$ and $\mathcal{L}[P] \to 1$ as they differ. We define the overall macroscale error as the average $\mathcal{L}_{\rm macro} = \frac{1}{3}\sum_{P} \mathcal{L}[P]$ of the errors computed for each distribution. 

We explore the 3-dimensional space of values for $k$, $\lambda$ and $\coupling$. To do so efficiently, we consider discrete values of $k$ from $k=1$ to $k=3$ in intervals of 0.05. We ensure that this is a range of plausible values for reproducing the atomistic measurements. For each $k$, we explore the 2-dimensional space of values for $\lambda$ and $\coupling > 0$, repeating each combination 48 times with a different realization of the initial conditions. Among the studied combinations of parameters, for each value of $k$ we consider only the pair $\lambda$ and $\chi$ with the lowest $\mathcal{L}_{\rm macro}$. We repeat this procedure for different mesoscale element lengths $l$. For each $l$, the size of the elasto-plastic lattice and the number $N$ of slip systems per element are set as detailed in Appendix \supsecref{system_sizes_etc}. We find that an excellent simultaneous agreement in all the compared macroscale magnitudes is possible for the different element lengths $l$ considered (see \figref{macro} for the specific case of $l=6.6$).

Moreover, a fit of similar quality is possible for a wide range of $k$ (\figref{optimization}a). Consequently, macroscale behavior can be regarded as a constraint that must be fulfilled by any acceptable set of elasto-plastic mesoscale parameters but does not provide us with enough information for the calibration of the model, for which knowledge of local mesoscale properties becomes mandatory.  

The error $\mathcal{L}[P(\extShearStressDrop)]$ in the fit to the distribution of external stress drops increases as the model resolution $l$ is reduced (\figref{optimization}a). External stress drops are the consequence of avalanches of slip events. The bigger the value of $l$, the bigger the portion that each mesoscale element covers, and therefore in this case avalanches become composed of fewer slip events. Eventually, for big enough $l$ avalanches become single events, with amplitudes that are fit as part of the model optimization. On the other hand, for smaller values of $l$, avalanches become collective events whose sizes are not explicitly part of the model optimization and emerge from the self-organization of plastic activity.

\subsection{Mesoscale local properties}
\label{sec:meso_properties}

For each simulated sample, an ensemble of patches representative of the system is defined, and for each patch, shear tests are performed with discrete orientations. Special attention is paid to the \emph{forward direction} $\orientationShear=0^{\circ}$ aligned with the external load, and the \emph{backward direction} $\orientationShear=90^{\circ}$, which will be used for the quantitative calibration of the mesoscale properties.

\begin{figure}
	\centering
	\includegraphics[width=12cm, keepaspectratio]{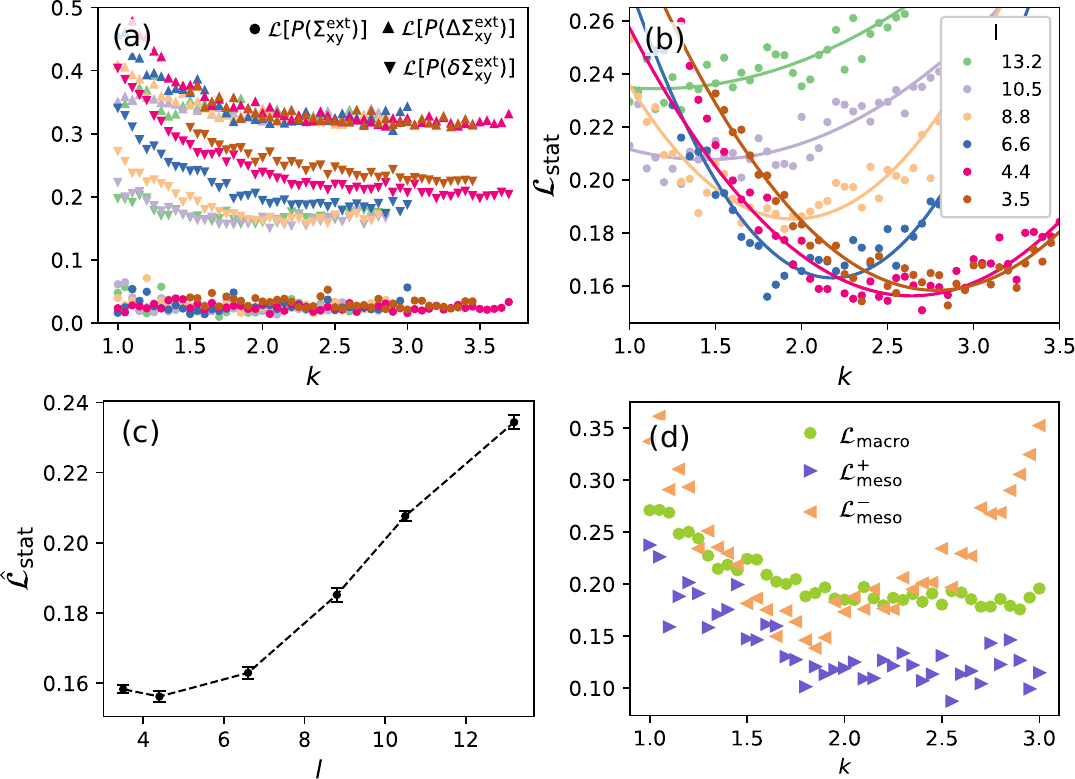}
	\caption{Relative errors in the fit as a function of the exponent $k$ of the Weibull distribution of slip thresholds for different mesoscale element lengths $l$ for: (a) the macroscale and (b) the aggregated macroscale and mesoscale properties. (c) Minimum of the aggregated error $\mathcal{L}_{\rm stat}$ as a function of $l$, where error bars indicate one standard deviation. (d) Individual components of $\mathcal{L}_{\rm stat}$ for $l=6.6$.}
	\label{fig:optimization}
\end{figure}

\begin{figure}
	\centering
	\includegraphics[width=15cm, keepaspectratio]{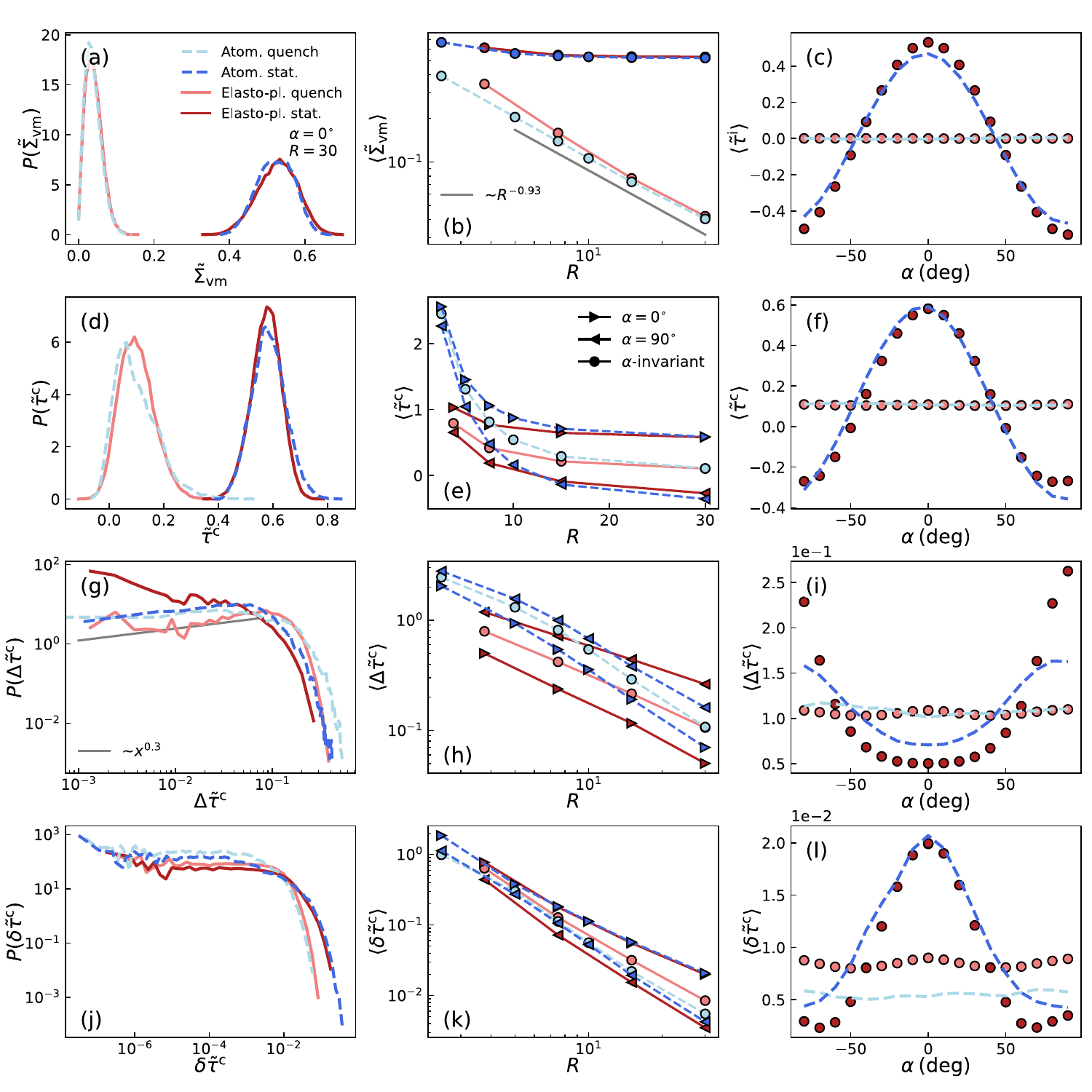}
	\caption{Comparison of the local properties of the atomistic (blue) and elasto-plastic (red) models in the quench and stationary regimes: von Mises stress $\PatchvonMisesStress$ and resolved shear stress $\PatchEffStress^{\rm i}$ on the shear test plane (first row), yield stress $\PatchThreshold$ (second row), distance to threshold $\PatchDisThreshold$ (third row) and stress drop $\PatchDropThreshold$ (forth row). The columns correspond to the probability distribution functions for $R=30$ (left), the averages as a function of $R$ (center) and the averages as a function of the shear orientation $\orientationShear$ for $R=30$ (right). The results shown correspond to the optimal parameters with $l=6.6$.}
	\label{fig:local_scales}
\end{figure}

We define a set of mesoscale magnitudes that allow us to quantitatively compare different aspects of the local mechanical response of both models when measured at the same scale $R$. Specifically, we consider the average and standard deviations of the local resolved shear stress $\PatchEffStress^{\rm i}(R,\orientationShear)$, yield stress $\PatchThreshold(R,\orientationShear)$, distance to threshold $\PatchDisThreshold(R,\orientationShear)$ and stress drop $\PatchDropThreshold(R,\orientationShear)$. By generically denoting the elasto-plastic measurements as $F_{\rm EP}(R,\orientationShear)$ and the atomistic ones as $F_{\rm MD}(R,\orientationShear)$, we define for these $8$ magnitudes the errors $\mathcal{L}[F]$ as 
\begin{equation}
\label{eq:meso_error}
\mathcal{L}[F] = \left| 1 - \frac{F_{\rm EP}(R,\orientationShear)}{F_{\rm MD}(R,\orientationShear)} \right|
\end{equation}
Although we will explore the impact of $R$ in the statistics of local properties, we compute the error functions using the largest available value of $R=30$, since presumably it is less affected by the rigid boundary bias introduced by the patch measuring method. We define the mesoscale error in the forward direction $\orientationShear=0^{\circ}$ as the average $\mathcal{L}^{+}_{\rm meso} = \frac{1}{8}\sum_{F} \mathcal{L}[F]$. We define the error $\mathcal{L}^{-}_{\rm meso}$ in the backward direction $\orientationShear=90^{\circ}$ in a similar fashion. The overall stationary state error, $\mathcal{L}_{\rm stat}$, is defined as the average between the macroscale and the mesoscale errors, $\mathcal{L}_{\rm stat} = \frac{1}{2}(\mathcal{L}_{\rm macro} + \frac{1}{2}(\mathcal{L}^{+}_{\rm meso}+\mathcal{L}^{-}_{\rm meso}) )$. We optimize $\mathcal{L}_{\rm stat}$ by exploring the neighborhood of the values that optimized $\mathcal{L}_{\rm macro}$ (\secref{macro_behavior}). This allows us to optimize the fit to the mesoscale properties while fulfilling the constraint imposed by the macroscale behavior.

As discussed previously, macroscale properties are not enough to establish the optimum set of mesoscale parameters. However, thanks to the addition of the mesocale properties, $\mathcal{L}_{\rm stat}$ exhibits a minimum $\hat{\mathcal{L}}_{\rm stat}$ at a specific value of $k$ (\figref{optimization}b). We fit a parabola $a+bx+cx^2$ to $\mathcal{L}_{\rm stat}$ to estimate the optimum value $\hat{k}$. Since to each value of $k$ we associate a pair of $\lambda$ and $\chi$, we can by interpolation obtain the optimum values $\hat{\lambda}$ and $\hat{\chi}$. As shown in \figref{optimization}c, $\hat{\mathcal{L}}_{\rm stat}$ depends on the element length $l$. We find an optimum value in the range of $l=4.4-6.6$. The error minimum is however rather shallow and the error starts to increase more appreciably only for larger element length. 

In order to further set the value of the element length $l$ we take into account several considerations: first, the model loses spatial resolution for the coarser element lengths (but remarkably can reproduce macroscale properties with excellent quality). Second, we also expect that the representation of STs requires a larger plastic strain amplitude for a lower element length $l$. Indeed, lowering $l$ leads to best-fit parameters with a decreasing value of $\coupling$, which implies larger local stress drops. We found that, below $l\approx5$, $\chi<1$. That is, the limit at which changing the sign of the resolved shear stress in the active slip plane becomes the dominant behavior. If this does not necessarily violate the non-negative dissipation criterion (\secref{slip_event}), it seems unlikely regarding the expected physics of STs. Third, the range $l=4.4-6.6$ is comparable to the lower limit established in \cite{Tsamados2009} for the validity of linear elasticity. Moreover, when employed as the patch size in the local yield stress method, it has been shown that this length scale optimizes the correlation between plastic activity and $\PatchDisThreshold$~\cite{Barbot2018}. Since we do not observe a significant difference in the error computed within that range, in the following, we consider the upper bound $l=6.6$. This scale is presumably above the typical size of atomistic rearrangements and larger than spatial correlations in the structural renewal process, both of which are fundamental assumptions of the elasto-plastic model.

We can have more insight into the origin of the error basin of $\mathcal{L}_{\rm stat}$ by looking at its components $\mathcal{L}_{\rm macro}$, $\mathcal{L}^{+}_{\rm meso}$ and $\mathcal{L}^{-}_{\rm meso}$ separately. As shown in \figref{optimization}d with $l=6.6$, the error $\mathcal{L}^{+}_{\rm meso}$ of the forward-oriented mesoscale properties exhibits qualitatively similar behavior as the macroscale error, with a wide range of parameters leading to quantitatively similar values without a clear minimum. However, the error $\mathcal{L}^{-}_{\rm meso}$ of the backward-oriented mesoscale properties exhibits a minimum at a specific combination of parameters. Consequently, among the parameters that reproduce the macroscale and the mesoscale properties aligned with the external load, only a specific combination $\hat{k}$, $\hat{\lambda}$ and $\hat{\coupling}$ can simultaneously reproduce the behavior of the patches when unloaded from the stationary state.

The estimated optimal parameter combination for the stationary state at the mesoscale $l=6.6$ is $\hat{k} = 2.18$, $\hat{\lambda} = 2.05$ and $\hat{\coupling} = 2.25$. The overall relative fit error $\mathcal{L}_{\rm stat}$ is 16\%. With the optimal parameters, we perform 1024 simulation runs. \figref{local_scales} shows von Mises stress $\PatchvonMisesStress =(\frac{1}{2}\PatchStress^{\prime}:\PatchStress^{\prime})^{1/2}$ (\figref{local_scales} first row), yield stress $\PatchThreshold$ (\figref{local_scales} second row), distance to the threshold $\PatchDisThreshold$ (\figref{local_scales} third row) and stress drop $\PatchDropThreshold$ (\figref{local_scales} fourth row) in the quench and stationary states. Their probability distributions and dependence on orientation measured with $R=30$ are shown in the first and third column, respectively. In the second column, we plot the averages measured for different scales $R$. The same analysis carried out for the standard deviations is reported in Appendix \supsecref{local_fluctuations}. A detailed overview of the fitting errors for the individual magnitudes is shown in \figref{relative_error}(model C).

We obtain a qualitative agreement in the quench and stationary states for all the investigated properties that progressively improves both on average and standard deviation as $R$ increases. At $R=30$, we find a good general quantitative agreement. In the quench state, the results are independent of $\orientationShear$ due to the lack of privileged orientation. On the other hand, we observe in the stationary state a dependence on shear orientation related to the mere static equilibrium ($\PatchEffStress^{\rm i}$) but also to the induced anisotropy ($\PatchThreshold$ and $\PatchDisThreshold$) known as the Bauschinger effect~\cite{Patinet2020}.

The local shear stress $\av{\PatchEffStress^{\rm i}}$ and the yield stress exhibit a remarkably good match for every orientation $\orientationShear$. However, the distances to threshold $\av{\PatchDeltaThreshold}$ and the stress drops $\av{\PatchDropThreshold}$ exhibit qualitative differences, suggesting that the elasto-plastic model captures the emergent anisotropy with a different degree of accuracy depending on the orientation $\orientationShear$. Specifically, we observe a systematically higher accuracy in the forward direction ($\orientationShear=0^{\circ}$) than in the backward direction ($\orientationShear=90^{\circ}$). Moreover, the distance to threshold in the backward direction exhibits an incompatible scaling with $R$. Another qualitative discrepancy between models are the small local minima (\figref{local_scales}l, \figref{local_fluctuations}d and f), presumably an artifact induced by the FEM quadrilateral structured mesh.

We note that \figref{local_scales}(e,f) reports negative yield stress $\av{\PatchThreshold} < 0$. Although slip systems have positive-definite slip thresholds $\SlipThreshold > 0$ the definition of coarse-grained yield stress (\secref{local_shear_tests}) allows the measurement of negative values. A detailed discussion can be found in Appendix \supsecref{negative_threshold}.

\section{Emergent properties}
\label{sec:emerging_properties}

The optimal elasto-plastic parameters are estimated by minimizing the discrepancies between models at the macroscale, and at a local scale only for the forward and backward orientations (see \secref{calibration_of_the_elasto_plastic_model}). We find that, for the optimal parameters found, a great deal of non-trivial and quantitatively accurate phenomenology naturally emerges without the need to explicitly include it into the optimization process.

\begin{figure}
	\centering
	\includegraphics[width=12cm, keepaspectratio]{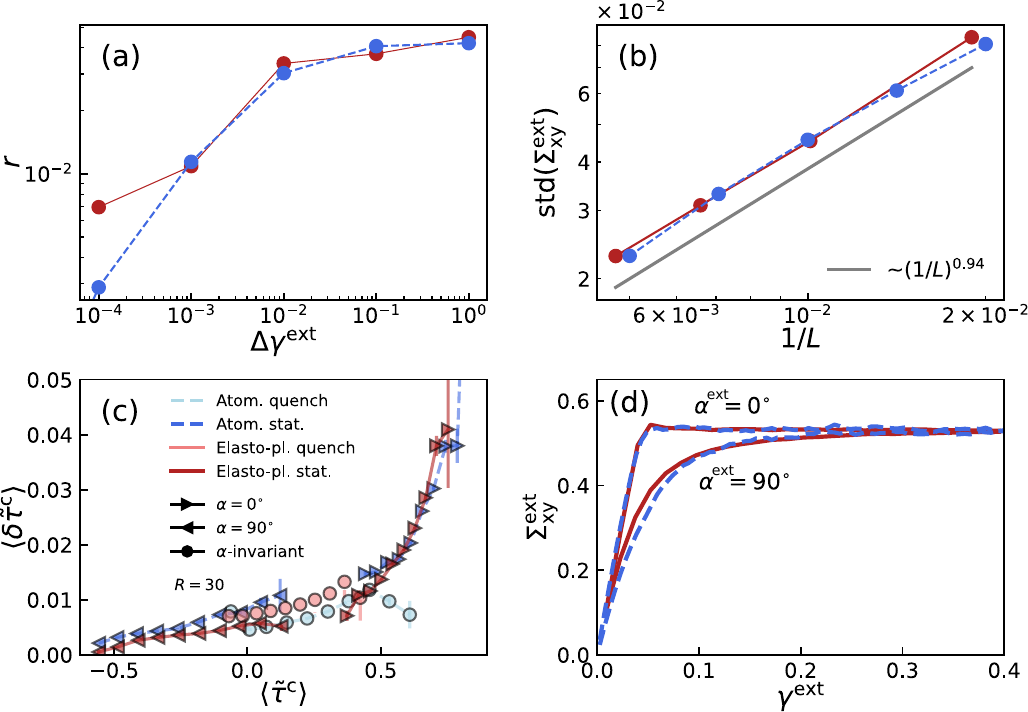} 
	\caption{Comparison of the behavior of the atomistic (blue) and elasto-plastic (red) models. (a) Root-mean-square deviation $r$ of the external stress. (b) Finite-size scaling of the external stress standard deviation $\textrm{std}(\extShearStress)$. (c) Average local stress drop versus average yield stress with $R=30$. (d) Bauschinger tests : reloading ($\extOrientationShear=0^{\circ}$) and reverse loading ($\extOrientationShear=90^{\circ}$) for systems previously unloaded from the stationary state. The results shown correspond to the optimal parameters with $l=6.6$.}
	\label{fig:emerging}
\end{figure}

An example of such finding is the root-mean-square deviation $r$ of the external stress and the system size dependence of stress fluctuation $\std{\extShearStress}$, reported in \figref{emerging}a and b, respectively. We compute the the root-mean-square deviation over external strain intervals as
\begin{equation}
\label{eq:roughenss}
r(\Delta\extShearStrain) = \langle \textrm{std}\left( \extShearStress |  \extShearStrain, \extShearStrain+\Delta\extShearStrain  \right)  \rangle_{\extShearStrain},
\end{equation}
where the average is performed over intervals of width $\Delta\extShearStrain$, centered at different positions $\extShearStrain$ of the stress-strain curve in the stationary regime. $r(\Delta\extShearStrain)$ increases and saturates to the $\std{\extShearStress}$, the latter decreasing with system size as $\sim L^{-0.94}$. We find that both stress correlation and standard deviation show excellent quantitative agreement with molecular statics simulations. 

Furthermore, in the elasto-plastic model specification we have defined a statistical relation between the amplitude of a slip event and the resolved shear stress $\SlipEffShear$ on the active slip system at the moment of activation (\Eqref{pdf_stress_drop}). To keep the model simple, we avoided a detailed description of the slip events and chose \Eqref{pdf_stress_drop} based solely on fulfilling the physical constraint of avoiding negative dissipation. Nonetheless, when coarse-grained stress drops $\PatchDropThreshold$ are measured, we observe an emergent non-linear relation between $\av{\PatchDropThreshold}$ and $\av{\PatchThreshold}$. This relation is shown in \figref{emerging}c for the quenched state and the stationary forward and backward directions, measured with $R=30$. The elasto-plastic model reproduces accurately the relation found in the atomistic system.

On the other hand, the elasto-plastic model consistently reproduces the anisotropy in the stationary state observed in the atomistic measurements for the different local properties studied, as discussed in \secref{meso_properties}. Note that such anisotropy is not a built-in ingredient of the model. Neither the laws of slip system renewal nor slip event performance consider a privileged orientation. Rather, the anisotropic response emerges naturally from the system's dynamics in the presence of the external load. This effect can be understood as the result of statistical hardening, i.e., the growth of the yield stress that occur due to a statistically biased slip activation. For simplicity, but without loss of generality, we consider the anisotropic response in terms of only forward and backward responses. In steady state, forward-oriented slip systems (i.e., $\orientationSlip \approx \extOrientationShear = 0^{\circ}$) experience a higher resolved shear stress. Consequently, forward-oriented slip systems suffer a statistical rise of slip threshold while backward-oriented (i.e., $\orientationSlip \approx \extOrientationShear + 90^\circ$) systems do not. 

To further assess this plasticity-induced anisotropy, we perform a reloading test for different orientations. To this end, systems are driven until the stationary state and from there are unloaded until they bear no external stress, $\extShearStress = 0$. While in the stationary state the behavior of the system is history-independent, once unloaded the future response depends strongly on the past deformation history, since the local yield stresses have been subjected to an anisotropic bias due to plastic deformation. When the unloaded systems are reloaded with $\extOrientationShear = 0^{\circ}$, we observe a nearly elastic-perfectly plastic response (\figref{emerging}d). However, when the system is loaded in the reverse direction $\extOrientationShear = 90^{\circ}$, we observe a much softer behavior showing a slow strain-hardening. The origin of this Bauschinger effect has been shown to come from the inversion of the polarization in the low distance to threshold population during unloading~\cite{Patinet2020}. As reported in \figref{emerging}d, we find a remarkable quantitative agreement between the atomistic measurements and the predictions of the elasto-plastic model regarding the plasticity-induced asymmetrical response to an external load.

\section{Successes, failures and key ingredients}
\label{sec:successes_failures_and_key_ingredients}

The elasto-plastic model reproduces all the phenomenology investigated in the stationary state with a relative aggregated error of $\mathcal{L}_{\rm stat} \approx 16\%$. Specifically, the model reproduces the macroscale behavior with an excellent agreement for all the element lengths $l$ investigated, despite the diminishing spatial resolution as $l$ increases. We find a good quantitative agreement at the mesoscale with $R=30$. However, when mesoscale properties are measured in the forward direction, i.e. aligned with the external load, such measures do not restrict the range of optimum mesoscale parameters with respect to the macroscale estimation. To further restrict the range, we measure the mesoscale properties oriented against the external load, which allows us to establish a specific combination of optimum mesoscale parameters. Moreover, the quenched state results generated by the Kinetic Monte Carlo method are of similar quality as the measures performed in the stationary state. These successes have been thoroughly discussed in the previous sections.

Despite the success of the elasto-plastic model, some disagreements are worth mentioning. We find that for the acceptable combinations of mesoscale parameters (i.e., that fulfill the constraint of matching the macroscale behavior), the elasto-plastic model systematically underestimates the distances to threshold in the forward direction in the stationary state (\figref{local_scales}h). Moreover, we find a wrong scaling in the left tail of the probability density, with a negative exponent in contrast with the marginal stability expected in elasto-plastic models and observed in our atomistic results. This discrepancy implies that, in the stationary state, elasto-plastic elements are closer to mechanical instability than atomistic regions are. We note that the non-coarse-grained single element response exhibits a scaling which is qualitatively right, with a positive exponent. 

On the other hand, measurements in the backward direction $\alpha= 90^\circ$ are, in general, less satisfactory than in the forward direction $\alpha=0^\circ$. A backward behavior with a lower fit quality is especially appreciable in the scaling of the distance to threshold $\av{\PatchDisThreshold}$ (\figref{local_scales}h). Since the distances to threshold and the forward-backward anisotropy result from self-organization, the mentioned discrepancies suggest that a new ingredient that modifies the dynamical evolution of the system might be required in the elasto-plastic description. 

A general trend in our results is the deviation of the elasto-plastic results from the atomistic ones at the lower scales. In the local properties, a convergence between measurements at coarser scales is expected since model-specific details and possible measuring artifacts related to local shear tests become less relevant. Nonetheless, this trend is also reflected in the macroscale behavior when looking at fine details. Specifically, \figref{macro}d reveals a discrepancy in the low-stress increment tail. While the elasto-plastic distribution exhibits discrete stress increments induced by the driving protocol, the atomistic distribution is smoothed. Such discrepancy in the low-stress increments results from elastic heterogeneity at the smallest scales, which leads to external stress increments that fluctuate around discrete values imposed by the discrete external strain increments. Moreover, the root-mean-square deviation of external stress (\figref{emerging}a) deviates below strain windows of $\Delta\extShearStrain < 10^{-3}$. This value is the lower bound above which both stress-strain curves exhibit a similar correlation structure between stress increments and drops.

Even before questioning the validity and the ingredients of the elasto-plastic model, other sources of error can be discussed, first of which is the accuracy of the atomistic local yield stress method itself. Since the method uses rigid boundary conditions, it is likely to overestimate the yield stress and underestimate the stress drops. This is in line with the underestimation of the local yield stress below $R<15$ in the elasto-plastic model, and consistent with the recent work done by Liu and coworkers~\cite{liu_elasto-plastic_2020}. Moreover, the measurement is purely local and does not take into account the elastic heterogeneities influencing the effective mechanical loading of each zone. All of these effects are expected to vanish asymptotically as the patch size increases, which is consistent with our results. These limits highlight the interest of developing flexible loading conditions for the atomistic local yield stress method, which will be the subject of future works.

\begin{figure}
	\centering
	\includegraphics[width=14cm, keepaspectratio]{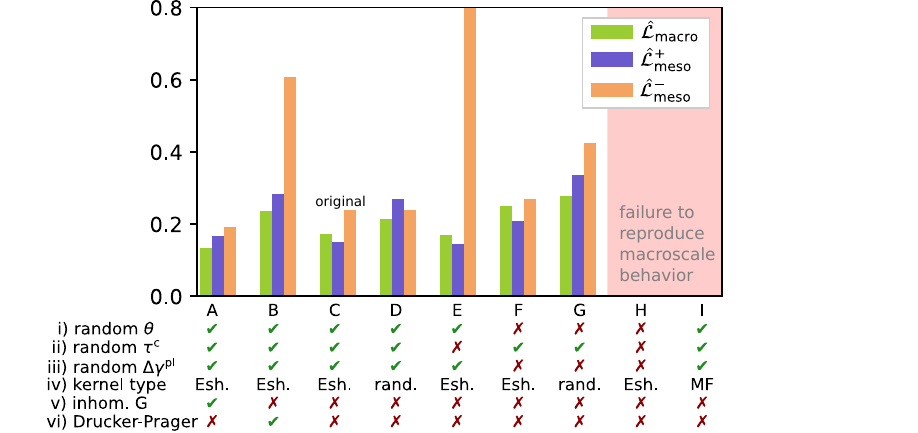} 
	\caption{Relative errors between the atomistic simulations and different versions of the elasto-plastic model (A-I) computed with their respective optimum parameters with $R=30$ and $l=6.6$ in the stationary state. Shown are the errors in the macroscale behavior ($\hat{\mathcal{L}}_{\rm macro}$) and mesoscale properties aligned with the external load ($\hat{\mathcal{L}}^{+}_{\rm meso}$) and against it ($\hat{\mathcal{L}}^{-}_{\rm meso}$). The different model ingredients are: (i) randomly oriented slip systems with their own threshold \emph{vs} scalar model with a single threshold value per element. (ii) Stochastic threshold renewal \emph{vs} fixed value. (iii) Stochastic slip amplitudes \emph{vs} fixed value. (iv) Eshelby, randomly shuffled Eshelby or mean-field interaction kernels. (v) Non-evolving and inhomogeneous \emph{vs} homogeneous shear modulus. (vi) Drucker-Prager \emph{vs} pressure-independent yield criterion.}
	\label{fig:model_comparison}
\end{figure}

In order to establish the impact of each ingredient employed in the elasto-plastic model, we finally quantify the different relative errors for various model flavors (see Appendix \supsecref{alternative_models} for a detailed description). As reported in \figref{model_comparison}, these errors are classified into three categories corresponding to macroscale and mesoscale properties in the forward and backward directions.

Elastic heterogeneity has been shown to become progressively more important for the correct representation of elasticity in glasses as the coarse-graining scale decreases~\cite{Tsamados2009}. Therefore, it is the first candidate ingredient to enhance the agreement between the magnitudes mentioned above and their respective atomistic measurements. As discussed before, elastic heterogeneity leads to small disagreements between the atomistic and the elasto-plastic model at the lowest scales. Thanks to our FEM computation of the elastic fields, we study the impact of inhomogeneous but non-evolving elastic properties and conclude that it has a small effect on the measurements performed on the elasto-plastic model and specifically does not improve systematic disagreements such as the distances to threshold.

In our mesoscale description, quenched elastic inhomogeneity leads only to subtle variations of the optimal parameters and a very slight improvement (\figref{model_comparison}A), but the global picture is not altered with respect to the homogeneous case. An explanation might be that the model is sensitive to very few fundamental ingredients, namely disorder in slip activation and amplitude of elastic coupling, while the physical origin of such ingredients does not alter the model dynamics. Since the leading effect of elastic inhomogeneity is an increment in the local stress fluctuations, the right fluctuation might already be captured, in an effective sense, by the calibrated homogeneous model. Along similar lines, a more accurate description based on evolving elastic properties induced by structural evolution might, through the correspondence between elastic inhomogeneity and eigenstrain~\cite{Lifeng2014}, have a similar impact as increasing the fluctuation of slip event amplitudes. Considering correlations between slip thresholds and elastic properties might also lead to further improvements in the model results.

We introduced pressure sensitivity by considering a Drucker-Prager-like criterion. As shown in \figref{model_comparison}B, the error increases with respect to the original model, especially in the backward direction. This finding is surprising since pressure sensitivity is known to be present in the atomistic system~\cite{Barbot2020}, and suggests that pressure effects are not correctly captured by a Drucker-Prager criterion alone as discussed in \cite{rottler_yield_2001,molnar_densification_2016}. An improvement would involve considering a local criterion for permanent local dilation and contraction based on free volume dynamics~\cite{Li2013}.

Along similar lines, we studied the impact of diverse extra model features that do not bring new fundamental ingredients. Namely, we considered (i) a fluctuating number $N$ of slip system per elasto-plastic element; (ii) a typical plastic strain amplitude controlling correlations between the previous and renewed thresholds; (iii) an explicit built-in anisotropy in the slip thresholds as a function of orientation, understood as a bias in the structural renewal process induced by external stress. These ingredients induce minimal changes in the overall picture presented in this paper, and we find that the mentioned deficiencies in the comparison between both models do not improve.

We finally follow the opposite approach by assessing the quality of simpler models. To this end, we implement a scalar model with only two slip systems of orientations $0^\circ$ and $90^\circ$, fixed slip amplitudes, and spatially homogeneous non-evolving slip thresholds (\figref{model_comparison}H). In this case, the only source of disorder is the quenched pre-stress field. Consequently, the stationary regime exhibits quasi-deterministic features, and the atomistic macroscale behavior cannot be reproduced accurately. If we restore the original stochastic renewal of the slip thresholds (\figref{model_comparison}F), a calibration of quality only slightly below the original model C is possible (note that this scalar model cannot, by construction, reproduce the behavior for all the orientations shown in the left columns of \figref{local_scales} and \figref{local_fluctuations}). However, if instead, we restored the stochastic slip orientations and amplitudes (\figref{model_comparison}E), the emergent plasticity-induced anisotropy is lost since the model fails to reproduce the backward response. Therefore, stochastic renewal of slip thresholds is the most physically relevant source of disorder in the system for reproducing the emergent anisotropy observed in the atomistic glasses.

The original model C uses an Eshelby-type elastic interaction obtained by solving the stress balance equation. In this case, slip events induce anisotropic stress variations with alternating signs as a function of orientation, which lead to spatial correlations in the stress field and to strain localization~\cite{Picard2004,Talamali2012}. To assess the role of spatial correlations, we consider a random interaction kernel obtained by element-wise randomly shuffling the stress variations induced by slip events in the rest of the system. This kernel maintains the same statistical properties as the original one except for spatial correlations. As shown by \figref{model_comparison}D, a calibration of quality only slightly below the original model C is possible. We take a step further and consider a homogeneous mean-field interaction by averaging the stress variations induced by slip events in the rest of the system (\figref{model_comparison}I). In this case, the atomistic macroscale behavior cannot be reproduced accurately. We conclude that stress fluctuations induced by slip events play a fundamental role in the steady state flow stress response, while spatial correlations do not. We note, however, that spatial correlations are mandatory to reproduce the strain localization observed in the transient regime observed in more stable glasses~\cite{Barbot2018}.

For all the model versions, the emergent relations depicted in \figref{emerging} remain qualitatively valid. The mean-field and the simplest model (\figref{model_comparison} I and H, respectively), however, fail to reproduce the root-mean-square deviation of the external stress (\figref{emerging}a) and thus the correlations in the stress-strain curve.

\section{Conclusions}
\label{sec:conclusions}

We have presented a comparison between two very distinct approaches to simulating the plastic deformation and structural evolution of glasses. On the one hand, we consider an atomistic model (\secref{atomistic_model_and_methods}), based on the detailed knowledge of individual atomic trajectories. On the other hand, we present an elasto-plastic model (\secref{elasto-plastic_model}), which coarse-grains details into a continuum mechanics description with local slip systems and random structural properties. We considered strain-driven systems in athermal quasistatic shear (AQS) conditions, which reduces the complexity of the process by, e.g., ruling out temperature and inertial effects, and therefore the number of ingredients necessary for a successful elasto-plastic simulation.

The method for measuring local properties proposed by \cite{Patinet2016} was extended here to an elasto-plastic implementation. We compared the results obtained with the same method but implemented with two different models describing the same system. Despite the differences in fundamental building-blocks and time and length scales, the method allowed us to compare the mechanical behavior independently of microscopic details. Thus, its applicability is wide and beyond atomistic simulations, and could be generalized to other models that can resolve microstructural details. Consequently, it is well-suited for the calibration of coarse-grained models based on data obtained at lower scales.

The elasto-plastic model reproduces the macroscale behavior in AQS conditions with a high degree of accuracy, although macroscale behavior alone is insufficient for determining a unique set of optimum mesoscale elasto-plastic parameters. To this end, we probed portions of the system and measured their mechanical response in isolation. By comparing observations at different scales with the atomistic system, we established the optimal values of the parameters. The model can reproduce the main phenomenology associated with plastic activity and structural evolution observed in the atomistic samples, such as the emergent anisotropy in the yield response induced by previous plastic deformation history. By calibrating the model, we have found an optimum mesoscale length within the range $l=4.4-6.6$. For coarser element lengths, the model loses spatial resolution but still reproduces macroscale behavior with an excellent quality.

Despite the generally good quantitative agreement between models, we have shown that some discrepancies at the local scales cannot be solved within the proposed model even for the coarser scales investigated. Moreover, simplified versions of the model are still able to reproduce the rich emergent phenomenology investigated. Therefore, it seems sensible that alternative novel ingredients are necessary to enhance the agreement between elasto-plastic descriptions and atomistic measurements at the mesoscale. Possible extra ingredients include using a more complex structural renewal process or the evolution of local elastic properties and their correlations with local yield thresholds and slip amplitudes. Moreover, the assumption of small deformations is only justified by model simplicity and computational performance, but is at odds with the large deformations undergone by the atomistic system when measurements are performed in the stationary state. Along similar lines, the consideration of advection~\cite{Nicolas2014} would capture material flows and rotations, a kind of structural evolution that is entirely missing in the presented elasto-plastic approach. Extra ingredients used in the literature that are missing in the present work are a distribution of element sizes and shapes in an unstructured mesh~\cite{Homer2009,karimi_role_2016,karimi_inertia_2017}, rounded potentials~\cite{jagla_smooth_potential_2017} for the activation of slip events, free volume creation/annihilation dynamics~\cite{Li2013}, the effects of viscosity and finite plastic relaxation times~\cite{liu_driving_rate_2016,liu_elasto-plastic_2020} or the effects of shear wave propagation~\cite{Puosi2014,Nicolas2015,karimi_role_2016}.

Along similar lines, future research is expected to be based on quantitative optimization of coarse-grained models that include diverse physically-motivated ingredients. Moreover, designing new optimization techniques, definitions of error between models, and improved methods for coarse-graining microscopic observations are natural steps forward to build upon the present work. The optimization of coarse-grained models through quantitative comparison with lower-scale reference data is a wide-open area, with the potential to impact the multi-scale modeling of mechanical properties of amorphous materials and enhance our understanding of the complexity emerging from the mechanisms at play.

\section*{Acknowledgements}
We are indebted to Damien Vandembroucq for an extended discussion and collaboration on this subject. D.F.C. and S.P. acknowledge the supports of French National Research Agency through the JCJC project PAMPAS (under grant ANR-17-CE30-0019-01) and the LabeX LaSIPS (under grant 20LL2000-129).

\bibliographystyle{crplain}

\nocite{*}

\bibliography{references_FRP_CRP_2020}

\appendix

\section{Appendix}
\label{sec:appendix}

\subsection{Impact of the number of slip systems}
\label{sec:number_slip_systems}

\begin{figure}
	\centering
	\includegraphics[width=8cm, keepaspectratio]{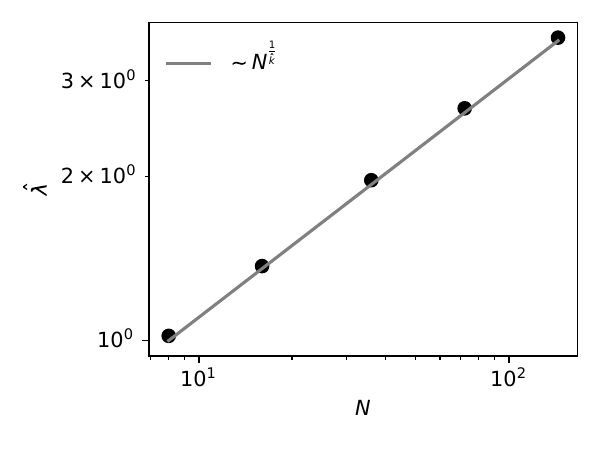} 
	\caption{ Optimal slip threshold scale parameter $\hat{\lambda}$ as a function of the number $N$ of slip systems in mesocale elements of length $l=6.6$.}
	\label{fig:lambda_vs_nslips}
\end{figure}

For a certain element length scale, we set a number $N$ of slip systems per element based on the density $\rho_{\rm s}$ of slip systems estimated in \secref{distribution_slips}. We assess the robustness of the results upon variation of the density $\rho_{\rm s}$. To this end, for a fixed scale $l=6.6$, we vary the number $N$ of slip systems per element and re-tune the optimum parameters $\hat{k}$, $\hat{\lambda}$ and $\hat{\chi}$. We find that the only effect is a rescaling of $\hat{\lambda}$ (\figref{lambda_vs_nslips}) according to the expected behavior for the Weibull distribution \Eqref{weibull_thresholds}. This behavior can be understood by noting that slip activation is controlled by the statistics of minimum distances to the threshold. If the number $N$ of slip systems present in an element increases, the minimum distance to the threshold in that element decreases. Therefore, to recover the typical distance to the threshold that best fits the atomistic data, the thresholds must be increased accordingly. The elasto-plastic results remain unaffected by variations of $N$, once the parameter $\hat{\lambda}$ has been properly rescaled.

\subsection{System sizes, number of slip systems and patch sizes}
\label{sec:system_sizes_etc}

In our analyses, we consider mesoscale elements of different length $l$. The physical domain represented by the elasto-plastic model has, in atomistic units, a side of length $L=98.8045$. To match the domain, we construct elasto-plastic lattices with a different number of elements $L_{\rm EP} \times L_{\rm EP}$. We compute the number of slip systems as $N=\rho_{\rm s}l^2$, where $\rho_{\rm s}=0.8$. However, we round $N$ to its closest value multiple of 4 to distribute the slip orientations as explained in \secref{distribution_slips}. Elasto-plastic patches are formed by $n \times n$ elements (\secref{local_shear_tests}), and each patch has an associated equivalent radius. Especially important is the patch with an equivalent radius of $R=30$, which we use for calibrating the elasto-plastic model. The values used for these parameters are reported in Tab.\ref{scales_details}.

\begin{table}
	\centering
	\begin{tabular}{c|c|c|c}
		$l$ & $L_{\rm EP}$ & $N$ & $n(R=30)$ \\ \hline
		3.5 &  29 & 8 & 15  \\ 
		4.4 &  23 & 16 & 12  \\ 
		6.6 &  15 & 36 & 8  \\ 
		8.8 &  11 & 64 & 6  \\ 
		10.5 &  9 & 92 & 5  \\ 
		13.2 &  8 & 144 & 4  \\ 
	\end{tabular}
	\caption{Element lenght $l$, elasto-plastic lattice size $L_{\rm EP}$, number $N$ of slip systems per element and elasto-plastic patch size $n$ with equivalent radius of $R=30$.}
	\label{scales_details}
\end{table}

\subsection{Initial configuration - Quenched super-cooled liquid}
\label{sec:supercooled_simulation}

Before applying the strain-driven protocol detailed in \secref{loading_protocol}, we must define the initial system configuration. To mimic the atomistic protocol as faithfully as possible, we consider an equilibrated super-cooled liquid (ESL) at a finite temperature $T$. The liquid is instantaneously quenched, and the resulting state is used as the initial system configuration. To this end, we simulate the atomistic rearrangements taking place in the ESL state within our elasto-plastic framework as a series of slip events that can spontaneously occur via thermal activation~\cite{Lemaitre2014}. Since we aim at a coarse-grained description of the process, we simulate thermal slip activation using the Kinetic Monte Carlo (KMC) method \cite{bulatov_stochastic_1994-2,Homer2009,Castellanos2018,Castellanos2019}.

The KMC method requires only knowledge of all the possible transitions to a next state and the energy cost associated with each one. Intermediate details of the transition process are not resolved. In the elasto-plastic model, a transition corresponds to the activation of a slip event among all the existing slip systems. Thus, one is probabilistically chosen from all the possible slip events based on the energy cost $\Delta E$ associated with their activation. We consider a thermal slip activation rate
\begin{equation}
\label{eq:total_kmc_rate}
\nu_{n} = \textrm{exp}\left( -\frac{\Delta E_{n}}{k_{\rm B} T} \right) 
\end{equation}
where the index $n$ refers to each slip system present in the system in a sequential manner, i.e., $n=1,2,...,NL^2$ where $N$ is the number of slip systems per element and $L^2$ is the number of elasto-plastic elements. We write the activation energy in terms of stress as $\Delta E_{n} = \Delta\SlipThreshold_n V_{\rm a}$, where $V_{\rm a}$ is a microscopic activation volume estimated to be $\approx 1$ from molecular dynamics. Assuming thermal activation as a Poisson process, the total activation rate is given by $\nu_{\rm tot} = \sum_{n=1}^{NL^2} \nu_{n}$. 

We aim to select a slip event with a probability proportional to its activation rate. To this end, first we compute the activation probability as $P_n = \nu_n/\nu_{\rm tot}$, from where we create a list of partial sums $S_n = \sum_{i=1}^{n} P_i$. Then, a random number $r$ from a uniform distribution in $(0,1]$ is drawn. We look in the list $S_n$ for the first index $k$ such that $S_k > r$. Slip event $k$ is then chosen by the KMC method. Since the probability for $r$ of landing in the portion of the partial sum $S_n$ is proportional to $\nu_n$, the algorithm chooses the slip event $n$ with probability $P_n$~\cite{Voter2007}. The chosen event is then performed (\secref{slip_event}), and the elastic fields are updated. After each thermally-activated event, an athermal adiabatic avalanche might be triggered, as described in \secref{loading_protocol}. For thermally-activated events, we set the maximum amplitude as $\gamma_{\rm max}(\SlipThreshold)$ (\Eqref{max_amplitude}), since the local stress $\SlipEffShear$ must reach, at the moment of activation, a value of at least $\SlipThreshold$ as result of thermal fluctuations.

Before starting the KMC simulation, the system is initially given random slip angles $\orientationSlip$ and thresholds $\SlipThreshold$ drawn from their respective distribution (see \secref{distribution_slips}). We consider no pre-stress $\preStress$ field and an initially homogeneous zero elastic strain $\elStrain=0$. Slip events are thermally activated under a fixed external strain $\extShearStrain=0$ until the slip thresholds, and the local stress field become statistically stationary, at which point the system has been equilibrated at temperature $T$. 

We mimic the atomistic instantaneous quench and subsequent relaxation by setting $T=0$ in the elasto-plastic system and performing an athermal mechanical relaxation, as described in \secref{loading_protocol}. The resulting state is used to create the initial configuration for the athermal quasistatic strain-driven protocol. Specifically, the local stress field of the quenched liquid is used as the pre-stress field $\Stress \to \preStress$ for the athermal solid, and the elastic strain is homogeneously re-set to zero $\elStrain=0$. Similarly, the slip angles and slip thresholds are used as the initial ones for the AQS simulation.

During the calibration process detailed in \secref{calibration_of_the_elasto_plastic_model}, we find that the properties of the stress field in the quenched state depend almost exclusively on the equilibration temperature $T$. On the other hand, the stationary state properties are independent of $T$, since they do not depend on the initial conditions. Consequently, we calibrate the value of $T$ in a straightforward manner based only on the quenched state stress field properties. For the optimal parameters reported in the main text, we find for $l=6.6$ a value $\hat{T}=0.25~V_{\rm a}$, a very reasonable value when compared with the molecular dynamics equilibration temperature $T=0.351$.

\subsection{Negative threshold}
\label{sec:negative_threshold}

We note that \figref{local_scales}(e,f) report negative average yield stress, $\av{\PatchThreshold} < 0$. Although slip systems have positive-definite slip thresholds $\SlipThreshold > 0$ the definition of coarse-grained yield stress $\PatchThreshold = \schmid(\orientationShear) : \critPatchStress$ (\secref{local_shear_tests}) allows the measurement of negative values. To understand this, let us consider a macroscale system in its stationary state, loaded along $\extOrientationShear=0^\circ$. If we unload the system until the first instability occurs, we need to perform, given its macroscopic size, a very small load increment with orientation $\extOrientationShear= 90^\circ$. Thus, when the instability occurs, the system's overall stress state remains oriented with $\extOrientationShear=0^\circ$. Consequently, the resolved shear stress at the moment of instability on the plane $\extOrientationShear= 90^\circ$ is negative, i.e., the yield stress is negative.

We can find a criterion for negative yield stress more rigorously by considering a shear test of orientation $\orientationShear$ performed on a patch belonging to a system at the stationary state. The initial resolved shear stress along the tested orientation is $\PatchEffStress^{\rm i}(\orientationShear) = \schmid(\orientationShear):\PatchStress^{\rm i}$. As a function of the von Mises stress $\PatchvonMisesStress =(\frac{1}{2}\PatchStress^{\prime}:\PatchStress^{\prime})^{1/2}$ it can be written as $\PatchEffStress^{\rm i}(\orientationShear) = \schmid(\orientationShear):\schmid(\PatchDevAngle):2\PatchvonMisesStress$. Taking into account that $\PatchDeltaThreshold = \PatchThreshold - \PatchEffStress^{\rm i}$ and $\schmid(\orientationShear):\schmid(\PatchDevAngle) = \textrm{cos}(2(\orientationShear-\PatchDevAngle))/2$, the condition for a negative yield stress $\PatchThreshold < 0$ is
\begin{equation}
\label{eq:condition_negative_threshold}
|\orientationShear - \PatchDevAngle| > \frac{1}{2} \textrm{arccos}\left(- \frac{\PatchDisThreshold(\orientationShear)}{\PatchvonMisesStress}  \right)
\end{equation}
Therefore, it becomes increasingly more likely to measure a negative yield stress $\PatchThreshold(\orientationShear) < 0$ the more the shear tests orientation $\orientationShear$ differs from the orientation $\PatchDevAngle$ of the patch initial state. On the other hand, a negative threshold becomes more likely to occur when the distance to threshold $\PatchDisThreshold(\orientationShear)$ is low compared to the patch's initial shear stress amplitude $\PatchvonMisesStress$. Moreover, since the distances to threshold decrease, on average, with $R$, negative thresholds are more likely measured at bigger scales.

\subsection{Local fluctuations}
\label{sec:local_fluctuations}

\begin{figure}
	\centering
	\includegraphics[width=10cm, keepaspectratio]{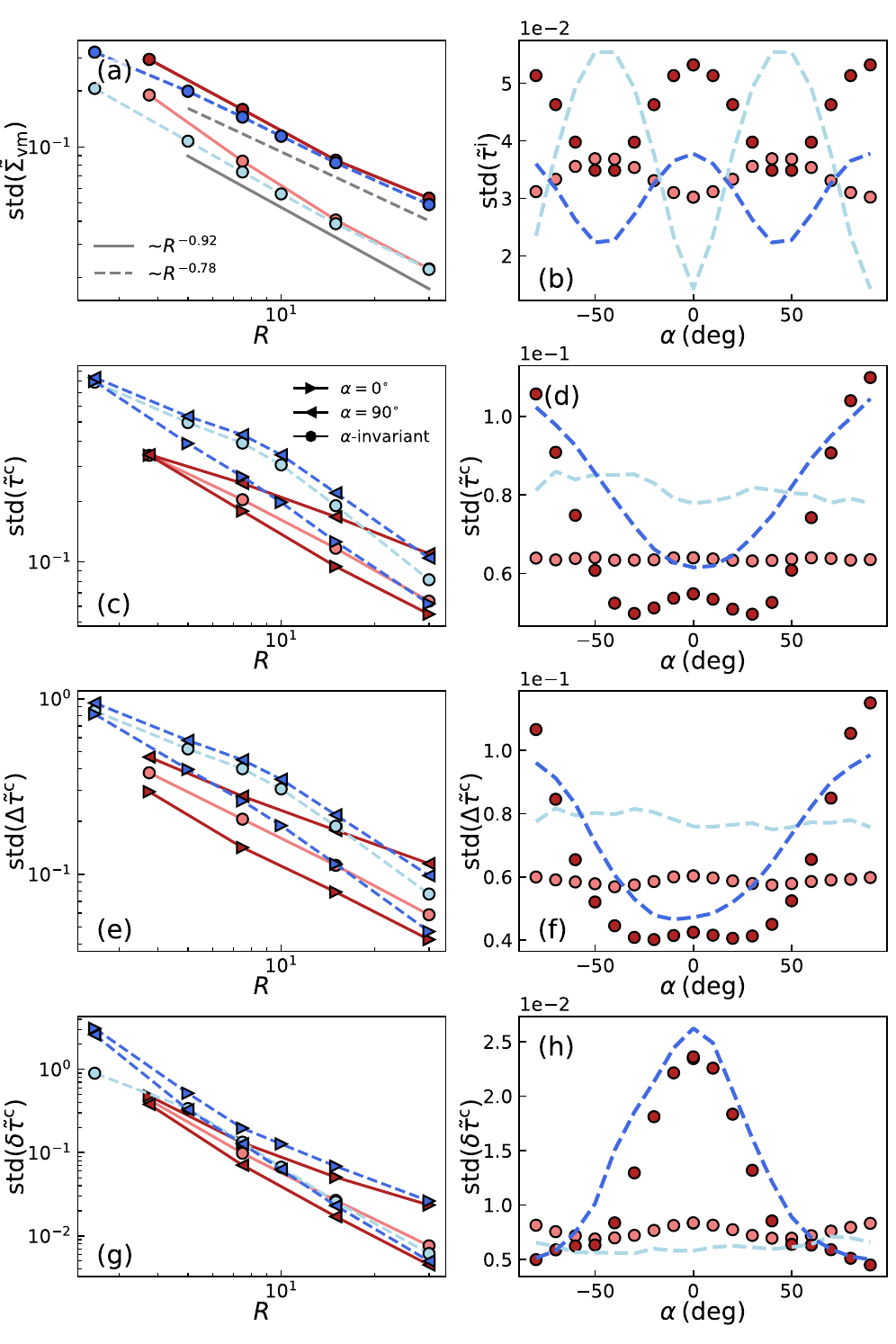} 
	\caption{Comparison of the local property standard deviations of the atomistic (blue) and elasto-plastic (red) models in the quench and stationary regimes: von Mises stress $\PatchvonMisesStress$ and resolved shear stress $\PatchEffStress^{\rm i}$ (first row), yield stress $\PatchThreshold$ (second row), distance to threshold $\PatchDeltaThreshold$ (third row) and stress drop $\PatchDropThreshold$ (forth row). The columns correspond to the averages as a function of $R$ (left) and the averages as a function of shear orientation $\orientationShear$ for $R=30$ (right). The results shown correspond to the optimal parameters with $l=6.6$.}
	\label{fig:local_fluctuations}
\end{figure}

We reach similar conclusions for the standard deviation of the explored local properties reported in \figref{local_fluctuations} as we did for the averages in the main text (see \figref{local_scales}). Namely, the elasto-plastic and atomistic model converge for big coarse-graining scales $R$. Nonetheless, we observed unexpected local minima in the vicinity of $\orientationShear \approx \pm 45^\circ$, which we attribute to artifacts induced by the FEM quadrilateral structured mesh. The quench state is successfully reproduced from our KMC approach.

\subsection{Backward local distributions}
\label{sec:backward_local_distributions}

\begin{figure}
	\centering
	\includegraphics[width=16cm, keepaspectratio]{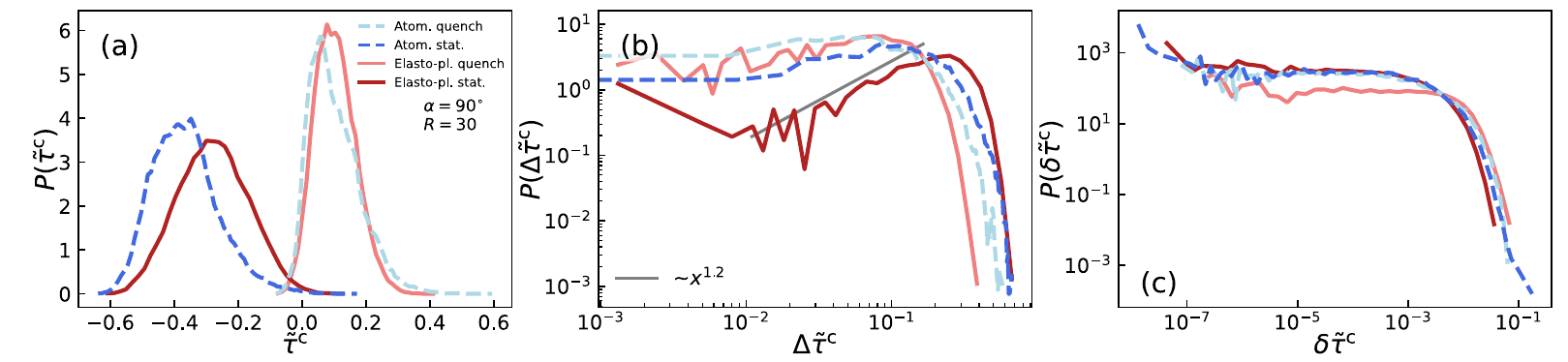} 
	\caption{Comparison of the probability density of the local stress drop for the backward direction $\alpha=90^{\circ}$ of the atomistic (blue) and elasto-plastic (red) models in the quench and stationary regimes for $R=30$: (a) local yield stress, (b) distance to threshold and (c) and stress drops. The results shown correspond to the optimal parameters with $l=6.6$.}
	\label{fig:local_backward_pdf}
\end{figure}

The probability densities of the explored local properties measured in the backward direction reported in \figref{local_backward_pdf} are in good agreement in the quenched state and a qualitative one in the stationary state. The local drops exhibit an excellent agreement in both cases.

\subsection{Limit to the amplitude of the slip events}
\label{sec:slip_amplitudes}

When a local slip event takes place an elasto-plactic element, a uniform eigenstrain is prescribed to the element. Therefore, a deforming element can be interpreted as an Eshelby inclusion with uniform eigenstrain embedded within an elastic matrix. As shown by \cite{vasoya_2020}, there exists a limit to the amount of eigenstrain an Eshelby inclusion can undergo to prevent negative dissipation. In the case of uniform eigenstrain, such limit is given by
\begin{equation}
\label{eq:vasoya_limit}
\boldmath{R}:\Strain^{\rm pl} \ge 0, \quad \boldmath{R} = (\Stress + \frac{1}{2}\Delta\Stress)
\end{equation}
where $\Stress$ is the local stress before the eigenstrain addition and $\Delta\Stress$ the change as a consequence of the addition. By means of the Eshelby tensor $\Eshelby$, defined by $\Delta\Strain = \Eshelby : \Delta \plStrain$, we can write the stress change as $\Delta\Stress = \Stiffness : (\Delta\Strain- \Delta\plStrain) = \Stiffness:(\Eshelby-\Identity): \Delta\plStrain$. Since we consider shear eigenstrain increments of the form $\Delta\plStrain = \Delta\plGamma \schmid$, with $\Delta\plGamma>0$, the criterion \Eqref{vasoya_limit} becomes
\begin{equation}
\label{eq:vasoya_limit2}
(\Stress + \frac{1}{2}\Stiffness:(\Eshelby-\Identity): \Delta\plGamma \schmid): \schmid \ge 0
\end{equation}
The factor $(\Stiffness:(\Eshelby-\Identity):\schmid):\schmid$ corresponds to the local shear stress variation induced by a local shear eigenstrain $\Delta\plGamma>0$ and is therefore negative. Rearranging \Eqref{vasoya_limit2} we arrive at the final form
\begin{equation}
\label{}
\Delta\plGamma \leq \frac{-2 \schmid:\Stress }{(\Stiffness:(\Eshelby-\Identity):\schmid):\schmid}
\end{equation}
To increase the numerical accuracy of our approach, we compute numerically the Eshelby tensor $\Eshelby$ of our quadrilateral finite elements. To do so, we prescribe local eigenstrains, solve the stress balance equation with the FEM and compute the induced strain response. The same operation carried out for different eigenstrain tensors allows us to obtain the components of $\Eshelby$ from a system of linear equations.

\subsection{Alternative models}
\label{sec:alternative_models}

\begin{figure}
	\centering
	\includegraphics[width=16cm, keepaspectratio]{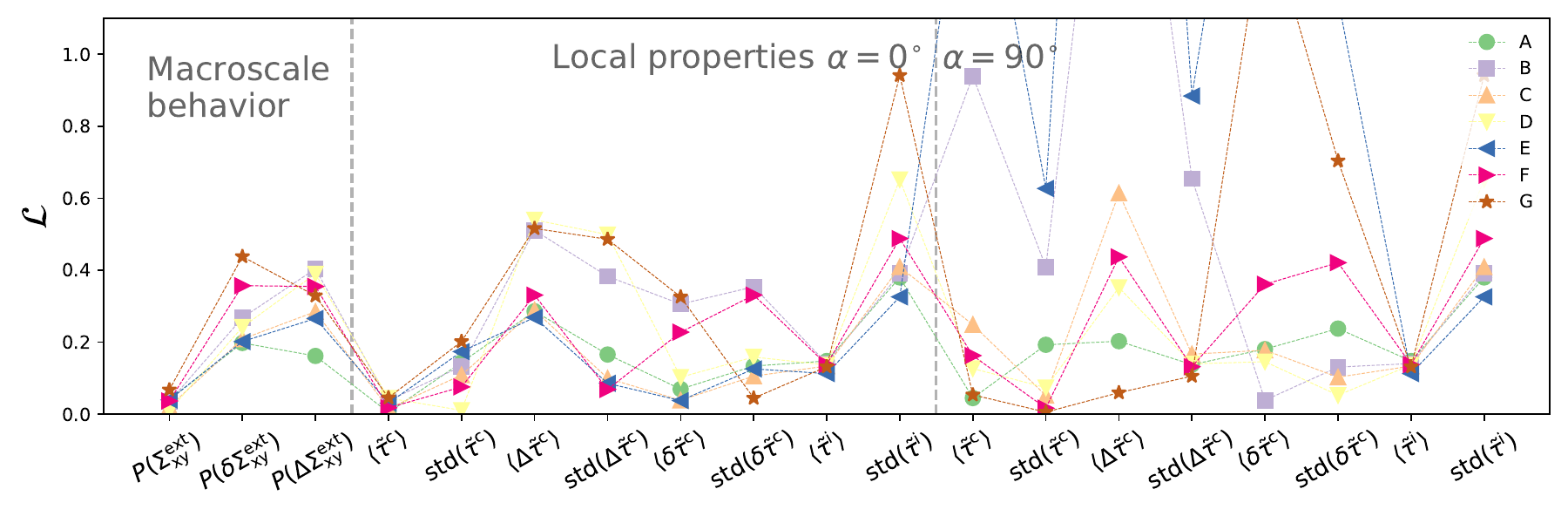} 
	\caption{Relative errors in the stationary state between the atomistic simulations and different versions of the elasto-plastic model computed with $R=30$ and $l=6.6$.}
	\label{fig:relative_error}
\end{figure}

We have quantified the fitting error of different models with $l=6.6$. Each model differs in certain implementation details and, therefore, in the optimal mesoscale parameters. This section details the differences between models and gives the values of the estimated optimum parameters. The fitting error for each compared quantity is reported in \figref{relative_error}.

\subsubsection{Model A}
\label{sec:inhom_shear_mod}

This model corresponds to the original model (\secref{elasto-plastic_model}) with spatially fluctuating, but non-evolving, values  of the shear modulus $G$, drawn from a Weibull distribution with average $\av{G}=18.4$ and $k_{G}=1.3$. The estimated optimal parameters are $\hat{k}=2.18$, $\hat{\lambda}=2.08$, $\hat{\chi}=2.19$.

\subsubsection{Model B}
\label{sec:press_sens}

This model corresponds to the original model (\secref{elasto-plastic_model}) with slip thresholds updated by the local pressure $p$ as $\SlipThreshold \to \SlipThreshold + \alpha p$, where $\alpha$ is the pressure sensitivity. Atomistic systems exhibit a non-zero average pressure $\av{p}=1.95$ with a sample to sample fluctuation of the average value of $\std{\av{p}}=0.22$. Since we do not consider a criterion for local free volume creation or annihilation, the average pressure cannot be reached dynamically. Instead, we set its value by manually adding a hydro-static contribution to the computed stress fields. The pressure sensitivity is set to $\alpha=0.2$~\cite{Barbot2020} by looking at the dependence of the yield stress $\PatchThreshold$ on pressure $\PatchPressure$, measured in patches with $R=30$. The estimated optimal parameters are $\hat{k}=1.88$, $\hat{\lambda}=1.05$ and $\hat{\chi}=3.33$.

\subsubsection{Model C}
\label{sec:original_model}

This model corresponds to the original model (\secref{elasto-plastic_model}), with optimum parameters $\hat{k} = 2.18$, $\hat{\lambda} = 2.05$ and $\hat{\coupling} = 2.25$.

\subsubsection{Model D}
\label{sec:orginal_rand_kernel}

We consider the original model with a randomized interaction kernel, which maintains the same statistical properties as the real Eshelby-type elastic interaction kernel but loses spatial correlations. To this end, we randomly shuffle the tensorial stress variations induced by yielding elements in the rest of the system. The local stress variation of the yielding elements is not shuffled. The estimated optimal parameters are $\hat{k}=2.03$, $\hat{\lambda}=2.53$ and $\chi=3.05$.

\subsubsection{Model E}
\label{sec:hom_thresholds}

We consider the original model with non-evolving homogeneous slip thresholds. The remaining sources of disorder are the heterogeneous pre-stress field, the stochastic strain amplitudes $\Delta\plGamma$, and the randomly oriented slip systems. In this case, we replace the free parameters $k$ and $\lambda$ by the slip threshold $\SlipThreshold$. The estimated optimal parameters are $\hat{\chi}=2.27$ and $\hat{\tau}^{\rm c}=0.925$.

\subsubsection{Model F}
\label{sec:minimal_model}

We consider a model with a simplified implementation with respect to the original one (\secref{elasto-plastic_model}). Namely, the elements contain only two slip systems of orientations $\orientationShear=0^{\circ}$ and $\orientationShear=90^{\circ}$, respectively. Both slip systems have the same slip threshold. Moreover, the consider constant slip amplitudes $\Delta\plGamma$. Consequently, we replace the free parameter $\chi$ by $\Delta\plGamma$. The estimated optimal parameters are $\hat{k}=1.17$, $\hat{\lambda}=0.38$ and $\Delta\hat{\gamma}^{\rm pl}=0.022$.

\subsubsection{Model G}
\label{sec:minimal_model_rand_kernel}

The same as \secref{minimal_model} but including the randomly shuffled kernel described in \secref{orginal_rand_kernel}. The estimated optimal parameters are $\hat{k}=1.09$, $\hat{\lambda}=0.57$ and $\Delta\hat{\gamma}^{\rm pl}=0.064$.

\subsubsection{Model H}
\label{sec:minimal_model_hom_thresholds}

The same as \secref{minimal_model} with non-evolving homogeneous thresholds as described in \secref{hom_thresholds}. No combination of free parameters leads to a satisfactory fit to the atomistic macroscale behavior.

\subsubsection{Model I}
\label{sec:MF_model}

We consider the original model with a mean-field homogeneous kernel. To this end, whenever a mesoscale element yields, we compute the average tensorial stress variation over all the non-yielding elements. The stress in all the non-yielding elements is updated with the computed average variation. No combination of free parameters leads to a satisfactory fit to the atomistic macroscale behavior.

\end{document}